\documentclass[twocolumn,prl,amsmath,amssymb,showpacs,superscriptaddress,floatfix]{revtex4}
\UseRawInputEncoding
\usepackage{graphicx}
\usepackage{bm}
\usepackage{lineno,hyperref}
\usepackage{epstopdf}
\usepackage{epsf}
\usepackage{xcolor}
\begin{document}
\title{Glass transition in monatomic systems: smearing of the same structure vs two structure competition}

\author{Yu. D. Fomin}
\affiliation{ Institute of High Pressure Physics RAS, 108840
Kaluzhskoe shosse, 14, Troitsk, Moscow, Russia}
\affiliation{Moscow Institute of Physics and Technology, 9
Institutskiy Lane, Dolgoprudny City, Moscow Region, Russia }

\author{E. N. Tsiok}
\affiliation{ Institute of High Pressure Physics RAS, 108840
Kaluzhskoe shosse, 14, Troitsk, Moscow, Russia}

\author{V. N. Ryzhov}
\affiliation{ Institute of High Pressure Physics RAS, 108840
Kaluzhskoe shosse, 14, Troitsk, Moscow, Russia}

\author{V. V. Brazhkin}
\affiliation{ Institute of High Pressure Physics RAS, 108840
Kaluzhskoe shosse, 14, Troitsk, Moscow, Russia}

\date{\today}

\begin{abstract}
In the present paper we discuss the properties of Voronoi polygons
in several monatomic glass- forming systems and compare them with
those of the Kob-Andersen mixture. We show that two mechanisms of
glass formation are possible: smearing of Voronoi polygons or
formation of polygons of two different shapes. Both mechanisms
lead to disturbance of the crystalline order in the system and
glass transition.
\end{abstract}


\maketitle

\section{Introduction}

Glasses are widely used in everyday life and in technology.
However, the microscopic mechanisms that give rise to this state
of matter remain the subject of controversy: glasses are
considered either as simply superviscous liquids, or as a result
of a true thermodynamic phase transition to a solid state
\cite{gr1,gr2,gr3,gr4,gr5,kobbinder,gr6,gr7,gr8,gr9,gr10,gr11,br,gg1,gg2}.
Glass is a general state of matter, and glass transition is common
in many-particle systems. Vitrification is observed on various
scales in corpuscular systems, ranging from colloidal suspensions
and granular materials to cell cultures. For this reason,
understanding this problem is so important to condensed matter
science.

In the most general sense, glass transition is a change in a
many-body system from an equilibrium liquid state to a
nonequilibrium disordered solid state. This change is not a phase
transition in the thermodynamic sense, at least not observed in
experiments, but is a kinetic phenomenon, in which an amorphous
solid does not have time to relax to an equilibrium state on
experimental time scales.

In ordinary condensed systems, the structure determines the
dynamics: for example, liquids are not ordered, while the
molecules move chaotically and relax to an equilibrium
thermodynamic state, that is, they lose their connection with the
initial conditions. At the same time, crystals are ordered and
have a nonzero static shear modulus, while, on average, the
positions of the molecules are retained over time. In this case,
changes in dynamic behavior follow from changes in the structure
caused by thermodynamic phase transitions, liquid - crystal. But
glass transition clearly does not fit into this paradigm.
Supercooled liquid slows down to a complete stop while maintaining
its liquid structure. In this regard, the main question in this
area is probably the question of whether the experimentally
observed glass transition is a purely dynamic phenomenon in which
the liquid stops kinetically, or the observed dynamics is a
consequence of the underlying phase transition from an equilibrium
liquid to the state of thermodynamic glass? In the first case,
glass is considered simply as a liquid with high viscosity that
grows indefinitely with decreasing temperature.

Glass transition temperature $T_g$ in this approach is a purely
conditional value corresponding to viscosity values of
$10^{12}--10^{13}$. In a liquid, poise and relaxation times of
$10--100 s$ are comparable to characteristic experimental times
(relaxation time corresponds to the average time between hops for
each particle). As an example of the dynamic theory of glass
transition, we can consider the dynamic facilitation theory
\cite{gg3}, associated with one of the main recent achievements in
understanding the physics of glass state, the discovery of the
phenomenon of dynamic heterogeneity \cite{gr12,gr13,gr14}. At the
moment, neither experiment nor modeling can give a definitive
answer as to whether glass transition is inherently a
thermodynamic or dynamic phenomenon, which opens up room for
theorists to develop new concepts. A deep connection between
static and dynamic descriptions is revealed by the theory of
random first-order transition (RFOT)
\cite{gr10,gr11,gr16,gr17,gr18}, based on the concept of existence
of an exponentially large number of metastable states at dynamic
transition temperature $T_d$ that is much higher than $T_g$, which
is the laboratory glass transition temperature.

At temperature $T_d$, a switch occurs from diffuse dynamics, in
which saddle points on the potential energy landscape dominate, to
activated dynamics, in which crossings of barriers dominate
\cite{gr14,gr15}. The behavior of the system at temperatures above
and below $T_d$ is described by the mode-coupling theory (MCT)
\cite{mct1,mct2} and RFOT, respectively. In experiments, the
crossover is often identified by the transition from Arrhenius to
superarrhenius behavior of relaxation times \cite{gr19,gr20}. At
temperatures $T> T_d$, transport is largely not collective, and
the topology of the state space is unremarkable. However, at $T
\rightarrow T_d$, the dynamics slows down and the system gets
stuck in a glassy metastable state. At $T <T_d$, there is a huge
number of statistically similar glassy metastable states separated
by barriers. Since there are so many glassy states, the system
will get stuck in one of the metastable glassy states. In the
absence of activated transport, it will remain in this state
forever. For this reason, $T_d$ is called a dynamic transition: it
is an abrupt transition only for the infinite range model, but in
general, it sets a temperature range, in which the dynamics
becomes glassy. In addition, at $T <T_d$, dynamic heterogeneity
plays an increasingly important role.

One of the facts which make it particularly difficult to study
glasses is that usually glass transition takes place in rather
complex substances such as network-forming liquids, metallic
alloys or organic liquids. At the same time, some works on
computer simulation give evidence that even monatomic systems can
experience glass transition (see, i.e., Refs.
\cite{glass88,dzugutov,we1,rysch}).

It is well known that the structure of a liquid does not undergo
very sharp changes during glass transition. The main structural
motif of glass transition is splitting of the second peak of the
radial distribution function $g(r)$ (RDF). However, even such
modest changes are taken into account within MCT and allow
predicting glass transition temperature.

A particular class of monatomic systems which can demonstrate
glass transition is systems with core-softened potentials. Glass
transition in different core-softened models was observed in a
number of publications \cite{glass88,dzugutov,we1,rysch}. In
particular, in Ref. \cite{rysch1} it was shown that there were
some special features of the RDFs of core-softened liquids, which
allow predicting their glass-forming ability.

Another class of systems, which demonstrate glass transition, is
systems with size polydispersity of particles (see, for instance,
Refs \cite{pd1,pd2,pd3}). It was shown in numerous articles that
polydisperse hard spheres \cite{pd1} or Lennard-Jones (LJ)
particles \cite{pd2,pd3} could demonstrate glass transition if the
degree of polydispersity was sufficiently large. At the same time,
a polydisperse LJ system looks very different from core-softened
systems and one can expect that it should vitrify via some other
mechanisms.

In the present paper we consider the peculiarities of the
microscopic structure of several monatomic systems, which
demonstrate glass transition, and compare them with those of a
glass-forming binary mixture (the Kob-Andersen (KA) mixture
\cite{ka1}). Based on these results, we show that two different
mechanisms are responsible for the failure of system
crystallization and its transformation into glass.

\section{Systems and Methods}

For this paper we simulated several glass-forming systems by means
of a molecular dynamics method. All simulations are performed in
the lammps simulation package \cite{lammps}. First, we simulated a
polydisperse LJ system, i.e., a system with continuous
distribution of atomic sizes (parameter $\sigma$ of the LJ
potential). The size of the particles was taken from the Gaussian
distribution with dispersion (the degree of polydispersity) up to
$\sigma=0.2$. A monodisperse LJ system was studied for the sake of
comparison. Second, we studied liquid silicon with the
Stillinger-Weber (SW) potential specially designed for simulation
of amorphous silicon \cite{vink}. The third system under
investigation was the so-called Repulsive Shoulder System (RSS)
\cite{we1}. Finally, we simulated a KA mixture, which is a
text-book system for investigation of glass transition \cite{ka1}.

In order to prove that the systems under investigation
demonstrated glass transition we calculated RDFs`a   ,
intermediate scattering functions (ISFs), and mean square
displacements (MSDs) (see Appendix A for the details of
calculations).

The key calculations of this work are related to the investigation
of the Voronoi cells of the systems. We calculated the volume of
the Voronoi cells and their surface. Using these parameters, we
propose a parameter which we call "sphericity":

\begin{equation}
  Sp=\frac{ (\frac{1}{4 \pi}S)^{3/2}}{\frac{3}{4 \pi}V}.
\end{equation}
In the case of an ideal sphere this parameter is equal to unity,
i.e., this parameter characterizes the deviation of the cell shape
from spherical. Although Voronoi cells cannot be spherical, the
different sphericities of cells mean that these cells have a
different shape, which is important for characterization of the
microscopic structure of liquid or glass. Apparently, the
structures with more nearest neighbors have Voronoi polygons with
more edges and approach a spherical shape closer than the
structures with fewer nearest neighbors. In particular, we
calculated the sphericity of an ideal face-centered cubic (FCC)
lattice and diamond structure. The corresponding sphericities are
$Sp_{FCC}=1.167$ and $Sp_{diam}=1.48$.

The sphericity parameter used in the present study is closely
related (but not equal) to the 'asphericity' parameter known from
the literature
\cite{ruocco1,ruocco2,shih,abascal1,abascal2,jedl1,jedl2}. The
main idea of both parameters is the same: they are equal to unity
for a sphere. However, we suppose that the sphericity parameter
defined in our paper is more convenient for calculations.

The results for amorphous silicon are expressed in physical units,
while the results for all other systems are given in dimensionless
units based on the parameters of the interaction potential (see
Appendix A for details).

\section{Results and Discussion}

\subsection{The Polydisperse Lennard-Jones system}

In this section we describe the results for the polydisperse LJ
particles. Five values of polydispersity are used: $\sigma=0.0$
(monodisperse particles), $\sigma=0.05$, $\sigma=0.1$,
$\sigma=0.15$ and $\sigma=0.2$.

\begin{figure}
\includegraphics[width=8cm, height=8cm]{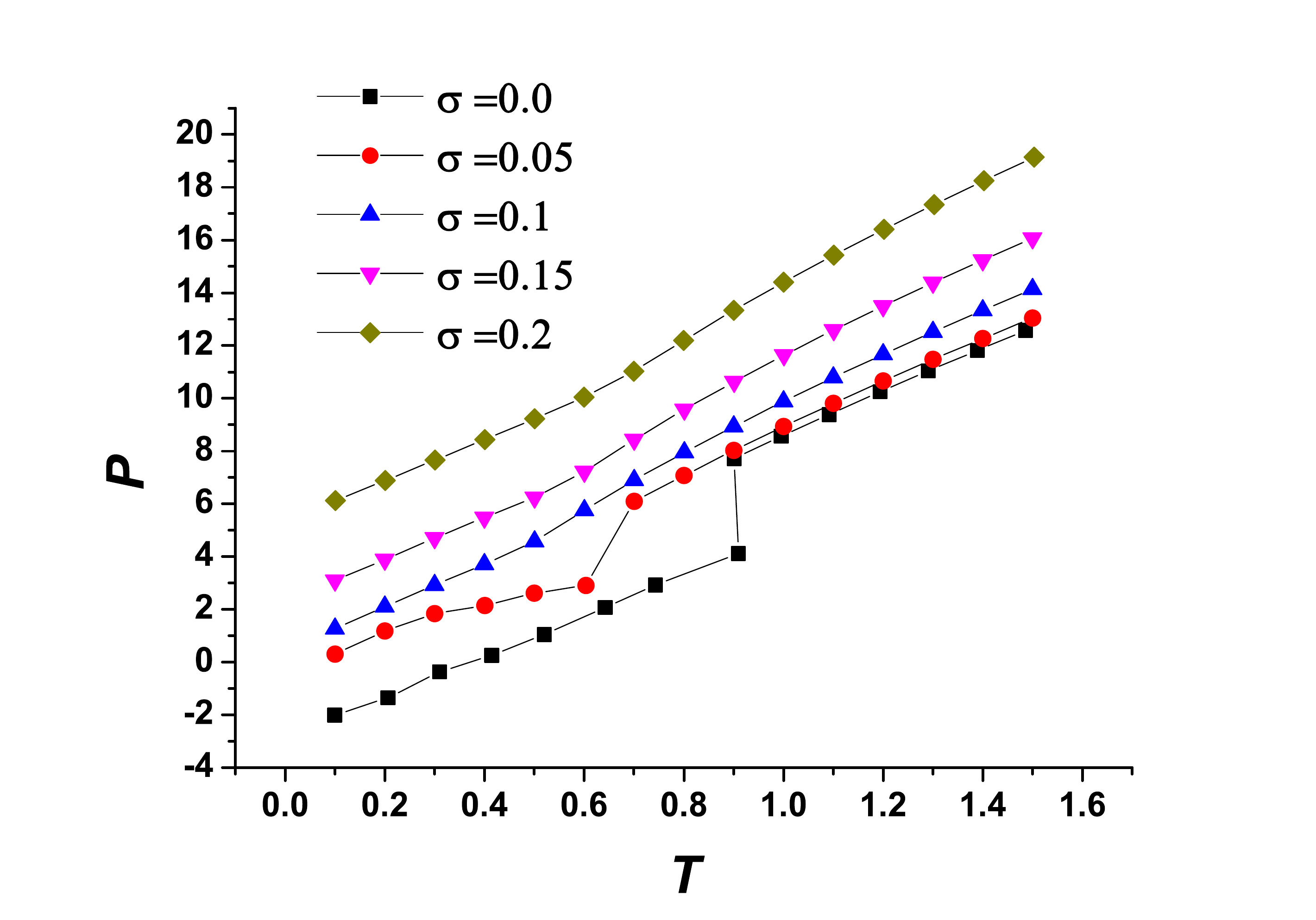}

\caption{\label{eos-pd} The equation of state of the polydisperse
LJ systems at different values of polydispersity. $\rho=1.0$.}
\end{figure}

In order to establish whether a system demonstrates a glass
transition we calculate the MSD $<r^2(t)>$ and ISFs
 $F_s(q,t)$ at different temperatures. The magnitude of the wave-vector is $q=14.04$,
which corresponds to the first maximum of the structure factor.
Figure \ref{eos-pd} shows the equation of state (EOS) of the
system for all values of polydispersity. One can see that in the
monodisperse system and at $\sigma = 0.05$ the EOS shows a sudden
bend which corresponds to the two-phase region of the first-order
phase transition. This part of the EOS coincides with the
corresponding part of the system phase diagram. At the same time,
at higher values of $\sigma$ the EOS demonstrates a smooth bend.
From this result we conclude that the monodisperse system and the
system with polydispersity $0.05$ crystallize while the systems
with higher polydispersity experience glass transition. This
conclusion is supported by the calculations of RDFs, MSD and ISFs
given in Appendix B.

\begin{figure}
\includegraphics[width=6cm, height=6cm]{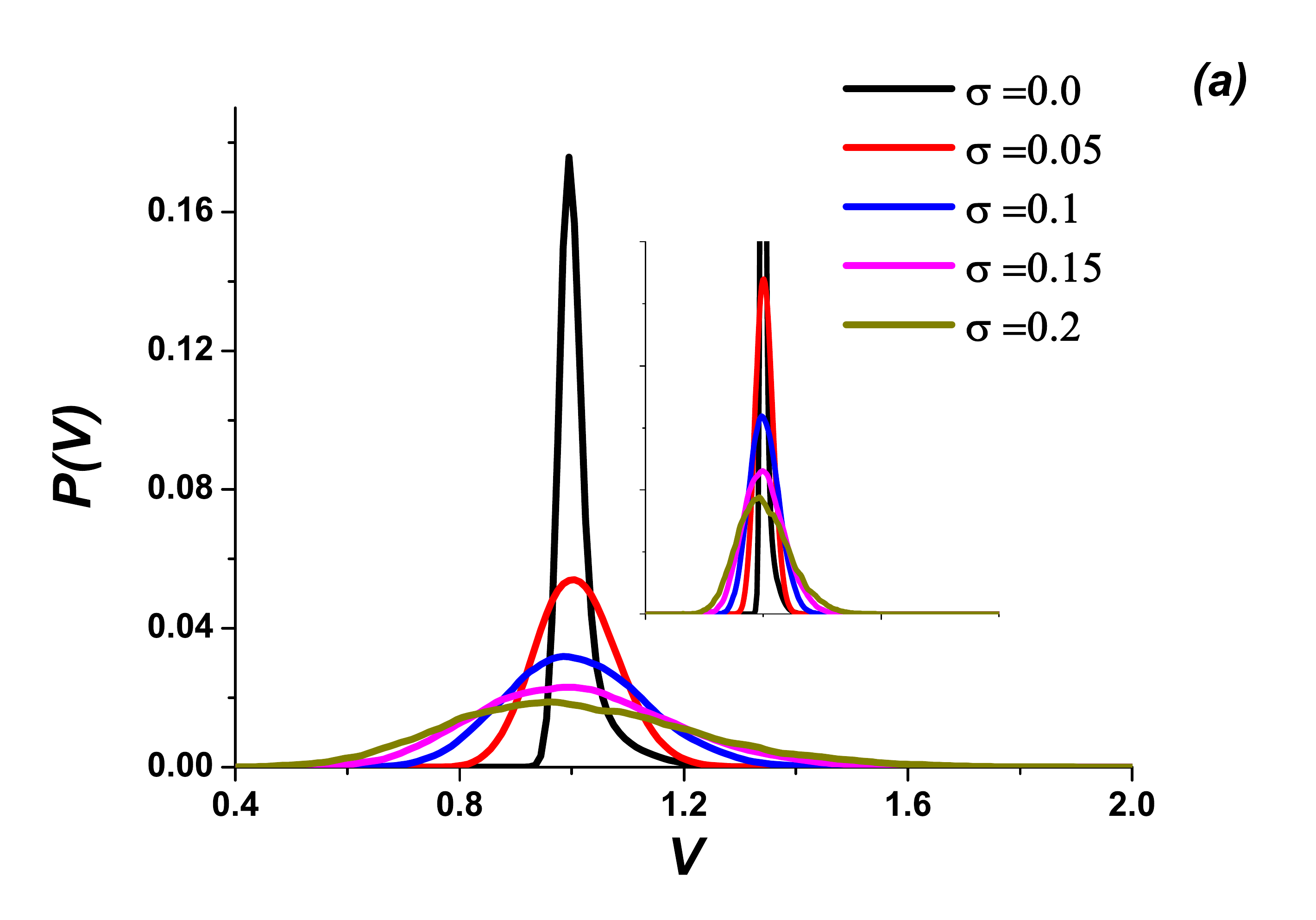}
\includegraphics[width=6cm, height=6cm]{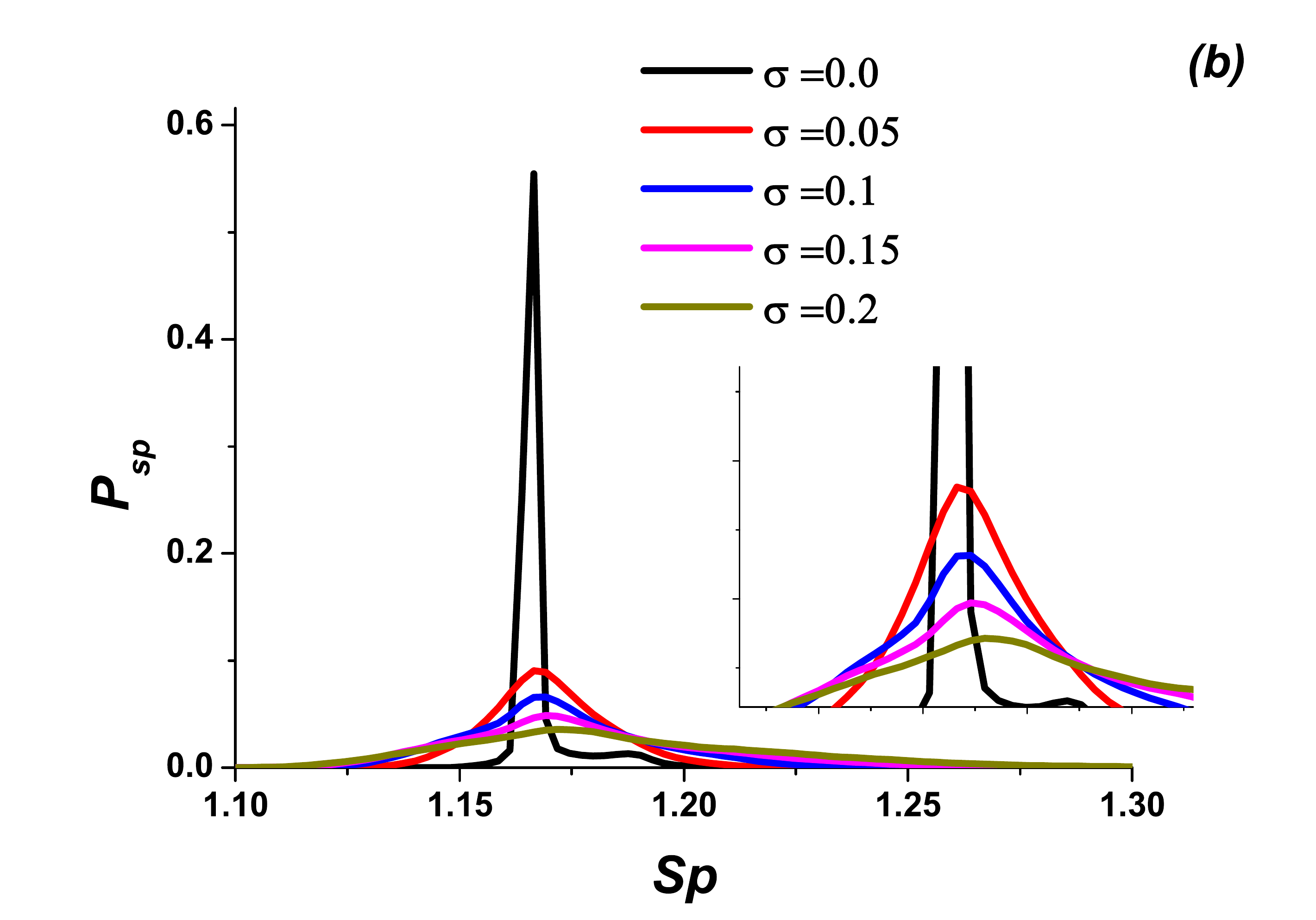}

\caption{\label{pv-pd} The distribution of the (a) volume and (b)
sphericity of the Voronoi cells of the polydisperse LJ system at
different values of polydispersity. $T=0.1$, $\rho=1.0$.}
\end{figure}

In order to see the difference in the local structure of these
five systems we analyze the properties of the Voronoi polygons at
the different values of polydispersity. In all cases we analyze
the configurations obtained at temperature $T=0.1$ which is far
below the crystallization or glass transition temperature. Figure
\ref{pv-pd} (a) shows the distribution of the volume of the
Voronoi cells for the systems with different polydispersity. In
the case of the monodisperse system the distribution demonstrates
a tall narrow peak. As polydispersity increases, the height of the
peak decreases and the peak becomes more spread. It corresponds to
the smearing of the local structure of the system.

Similar conclusions can be made from the sphericity of the Voronoi
cells shown in Fig. \ref{pv-pd} (b). It also shows a tall narrow
peak in the case of the monodisperse system, which becomes spread
in the polydisperse ones.

From this observation we conclude that in the case of the
monodisperse LJ system all particles are surrounded by Voronoi
cells of similar size and shape, which allows constructing a
crystal. At the same time, in the polydisperse systems the volume
and shape of the Voronoi cells become widely spread, which
prevents the formation of a regular lattice and serves as a
mechanism of glass transition in this type of systems.

\subsection{Amorphous Silicon}

A mechanism responsible for glass transition in the polydisperse
LJ system is described above. In order to show that the same
mechanism can also take place in other systems we describe here
the results for amorphous silicon.

We simulate the behavior of amorphous silicon by the SW potential
with the parametrization from Ref.  \cite{vink}, which is
specially designed for amorphous silicon. In Ref. \cite{vink} it
was shown that this potential gave a reasonably good description
of the properties of amorphous silicon. At the same time, this
model strongly overestimates the melting temperature. We find that
at $T=2000$ K the system spontaneously crystallizes into a diamond
lattice. Because of this we do not construct an EOS of this
system, but simulate an amorphous state at $T=800$ K, which should
be properly described by the employed model.

The RDFs, MSD and ISFs of this system are given in Appendix B.
Figures \ref{pv-si} (a) and (b) show the distribution of the
volumes and sphericities of the Voronoi cells for amorphous
silicon at $T=800$ K. One can see that these figures are
qualitatively identical to the ones of the polydisperse LJ system,
i.e., a spread low peak is observed. Therefore, glass transition
in amorphous silicon is related to the smearing of the local
structure of particles.

\begin{figure}
\includegraphics[width=8cm, height=8cm]{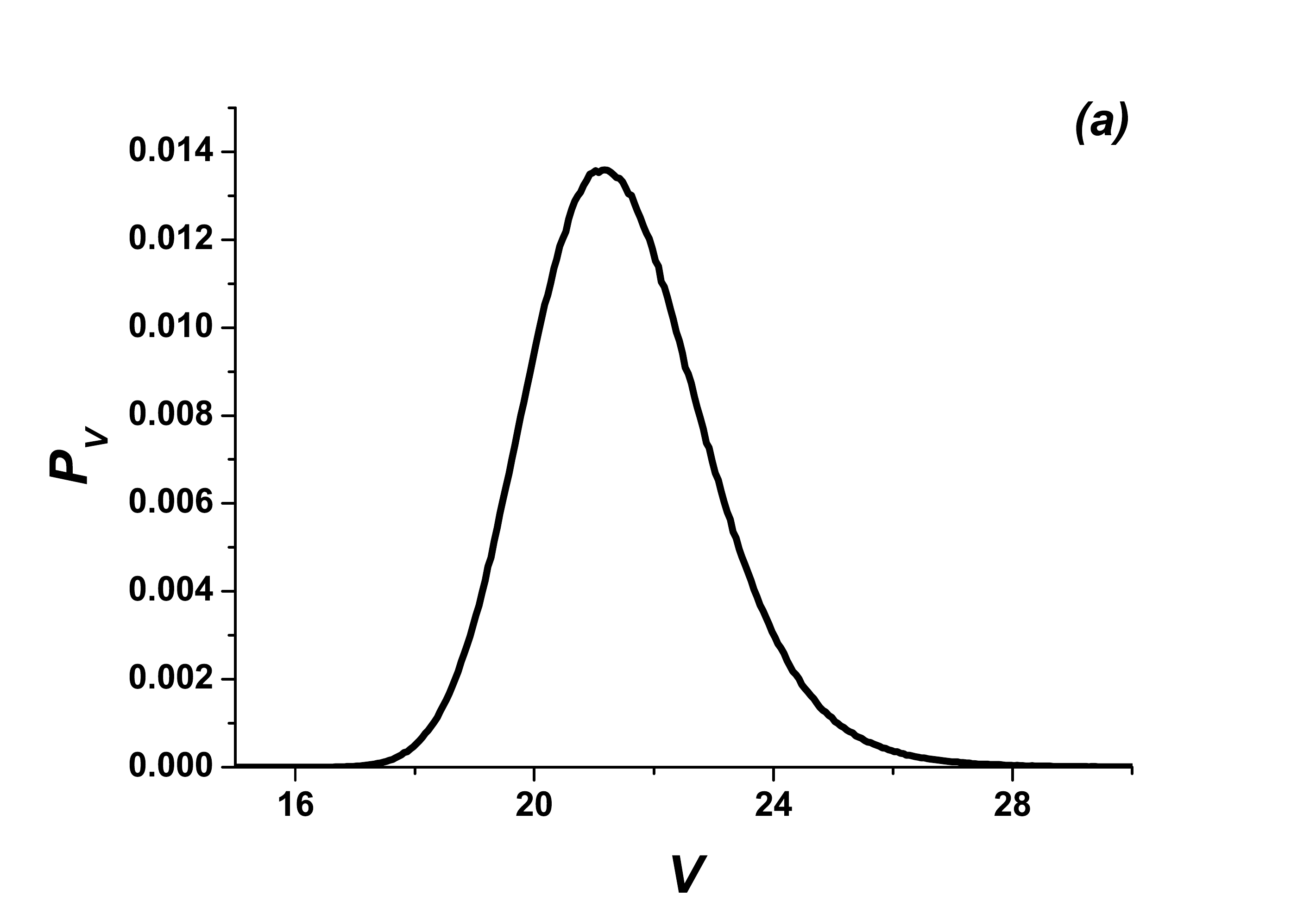}
\includegraphics[width=8cm, height=8cm]{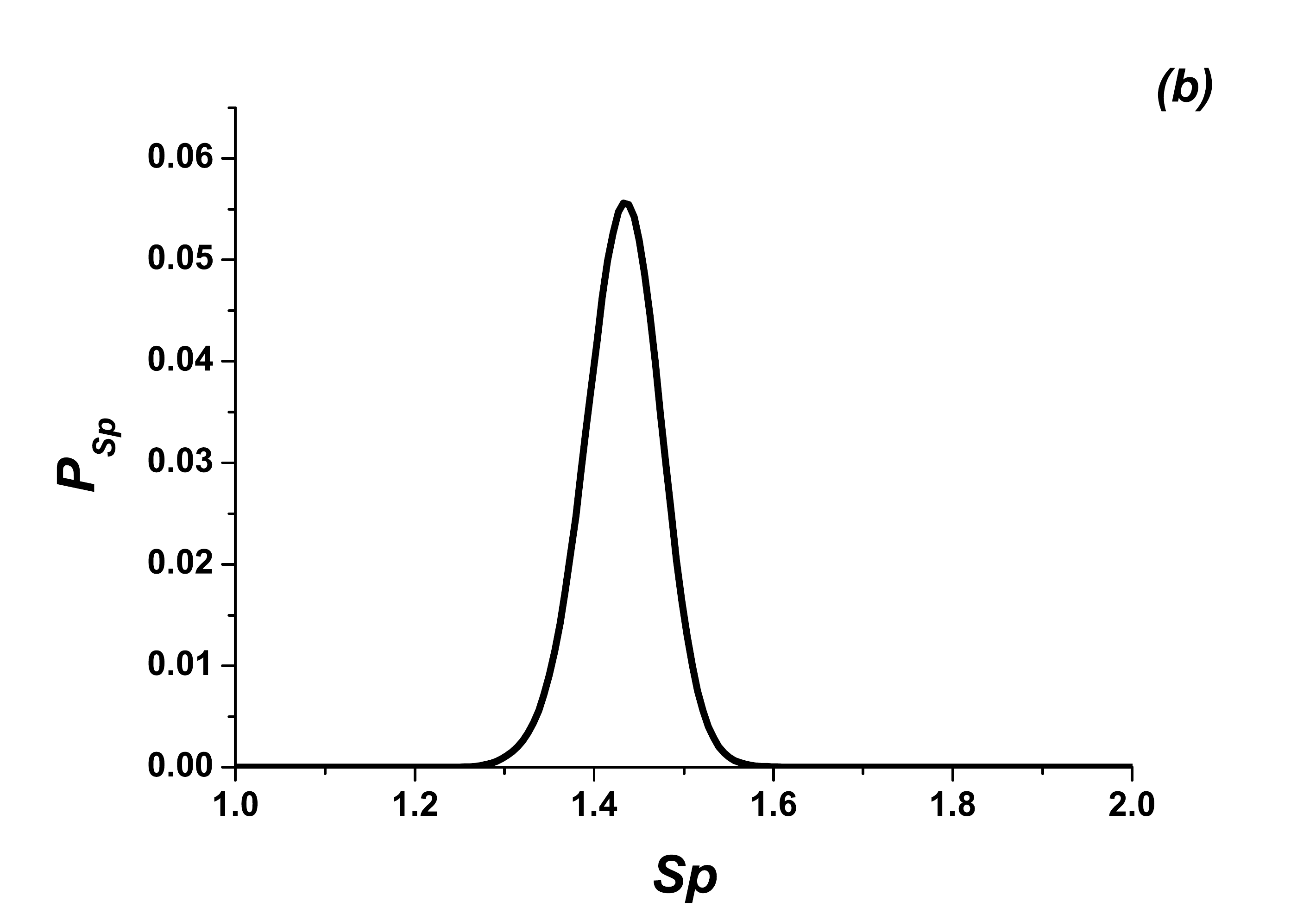}

\caption{\label{pv-si} The distribution of the (a) volume and (b)
sphericity of the Voronoi cells of amorphous silicon. $T=800$ K.}
\end{figure}

\subsection{The Repulsive Shoulder System}

Another mechanism of glass transition is found in the RSS. Glass
transition in this system is described in Refs.
\cite{we1,rysch,qcgl}. In the present work we also recalculate the
MSD and ISFs of this system. The results are given in Appendix B.

Figure \ref{pv-step} shows the distribution of volumes and
sphericities of the Voronoi cells of the RSS. One can see that the
distribution of volume has a single peak, which, however,
demonstrates two shoulders from both left and right sides. At the
same time, the distribution of sphericities demonstrates two
peaks, i.e., the Voronoi cells are characterized by two different
shapes. This result is in agreement with the quasi-binary nature
of the system induced by the shape of the interaction potential
\cite{we1}. Therefore, in the case of the RSS glass transition is
induced by a competition of two local structures, which have
similar volumes but different shapes. Below we show that the same
mechanism is responsible for glass transition in genuine binary
mixtures.

\begin{figure}
\includegraphics[width=8cm, height=8cm]{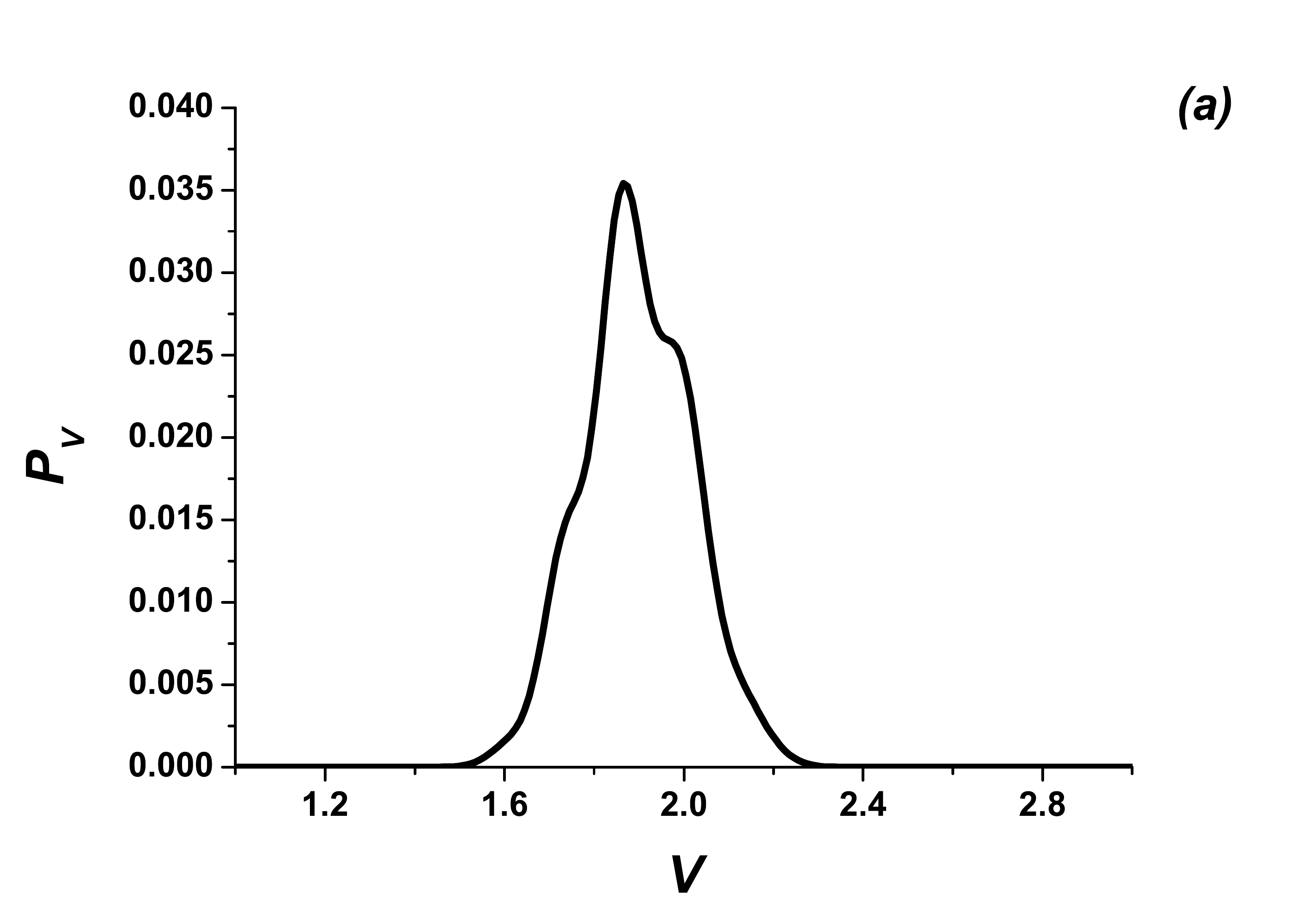}
\includegraphics[width=8cm, height=8cm]{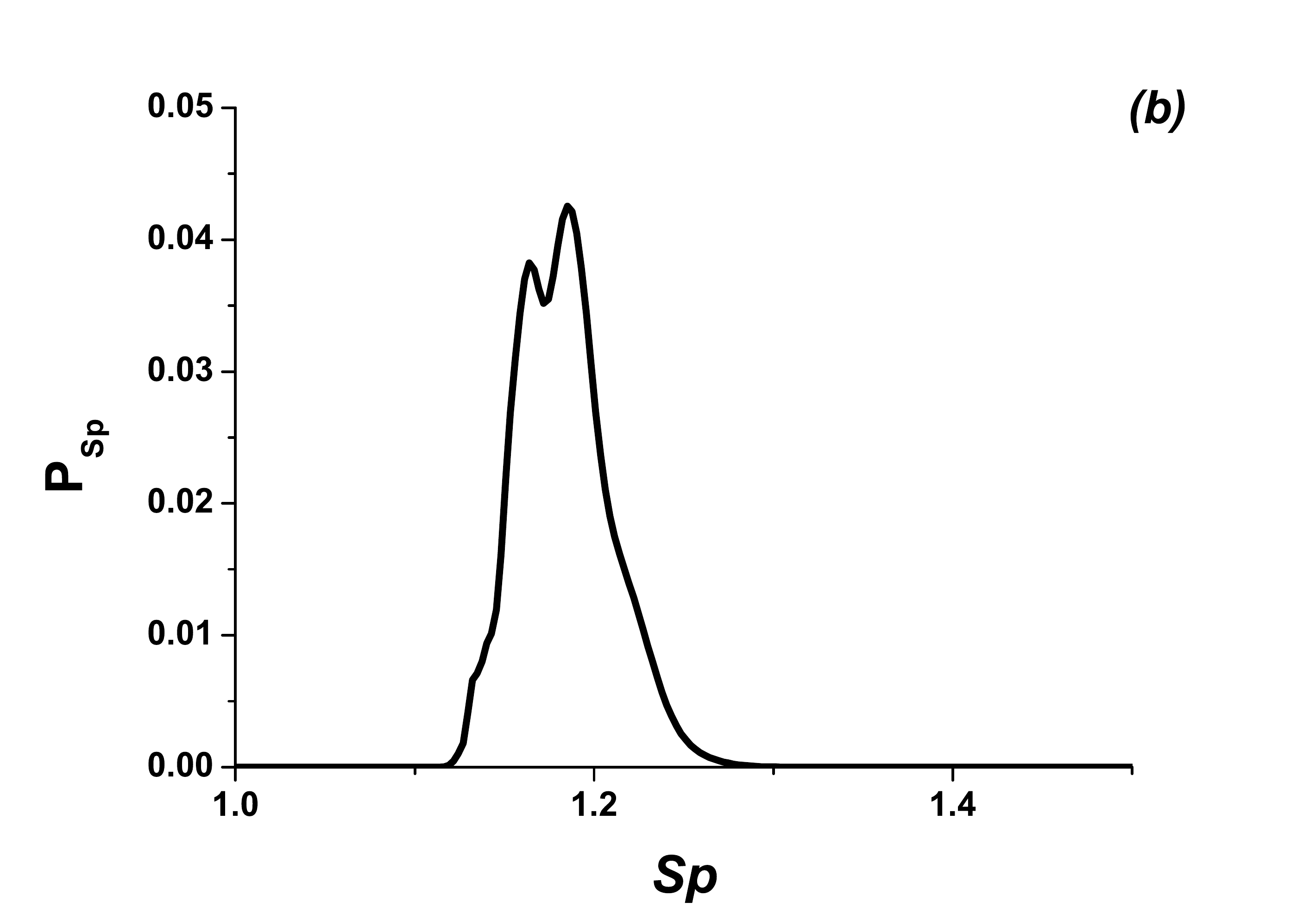}

\caption{\label{pv-step} The distribution of the (a) volume and
(b) sphericity of the Voronoi cells of the RSS. $T=0.06$,
$\rho=0.53$.}
\end{figure}

\subsection{The Kob-Andersen mixture}

\begin{figure}
\includegraphics[width=8cm, height=8cm]{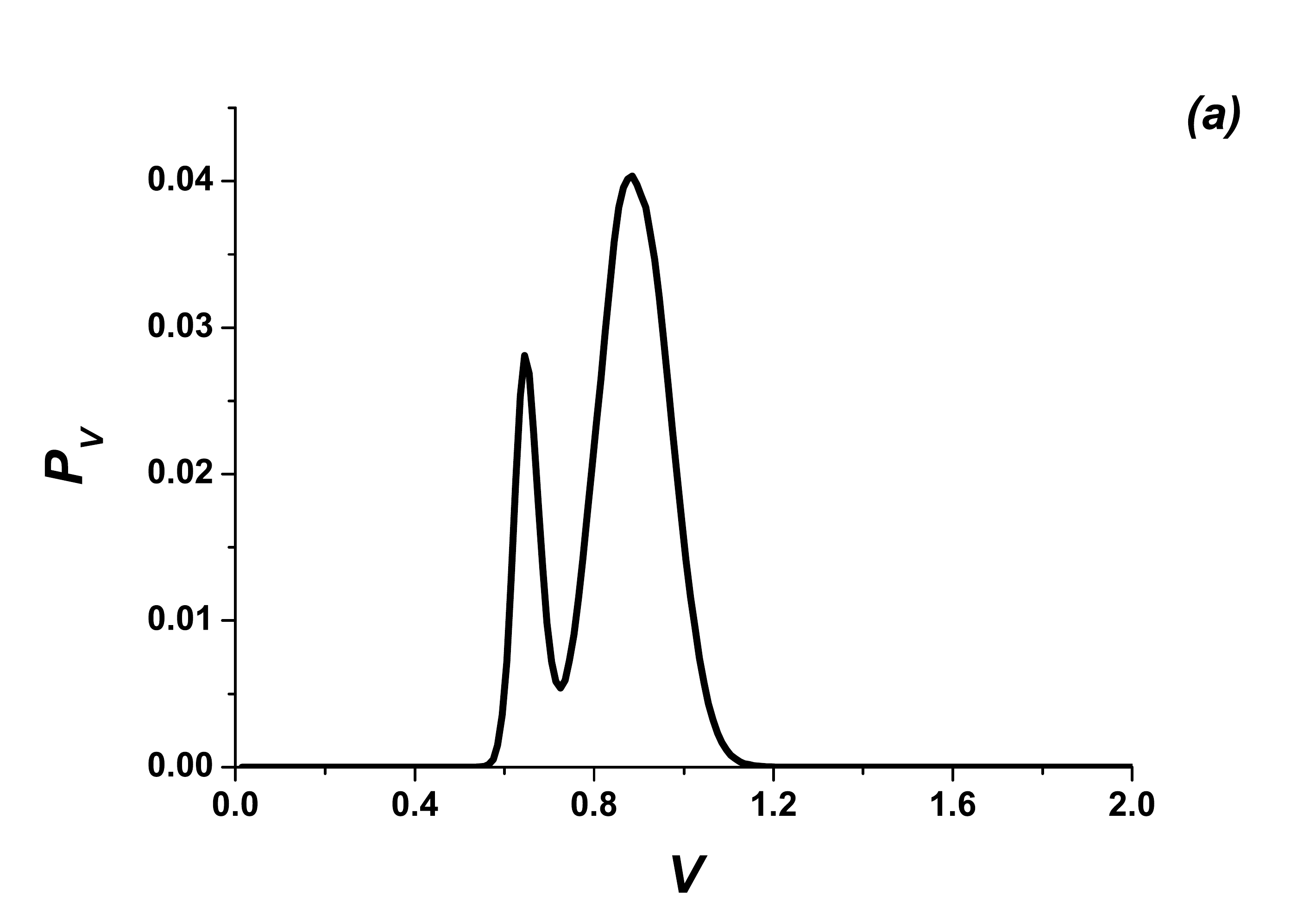}
\includegraphics[width=8cm, height=8cm]{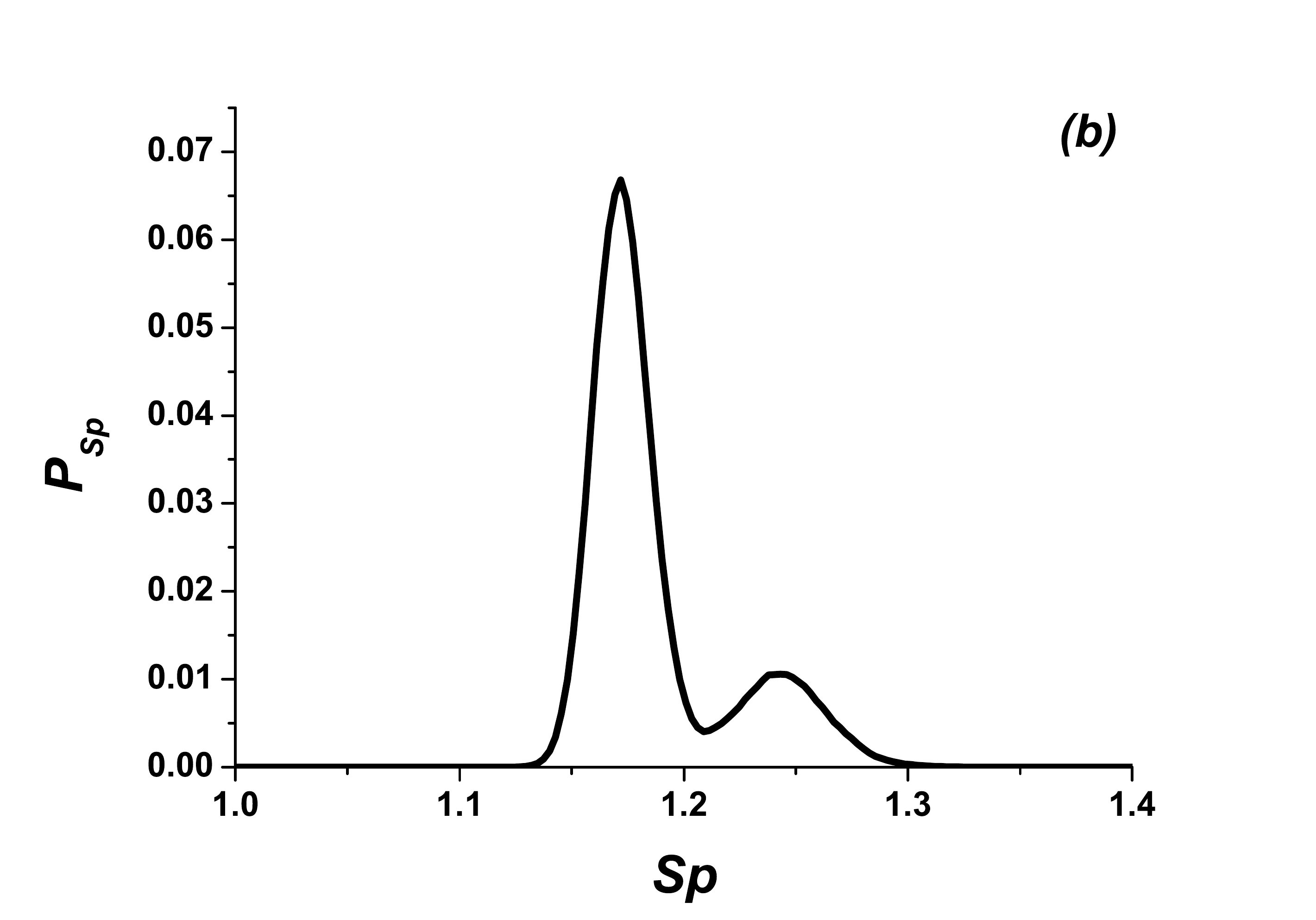}

\caption{\label{pv-ka} The distribution of the (a) volume and (b)
sphericity of the Voronoi cells of the KA mixture. $T=0.1$,
$\rho=1.2$.}
\end{figure}

In this section we describe the results for the KA mixture, which
is a text-book example of glass-forming liquids. Figure
\ref{pv-ka} shows the probability distribution of the volumes and
sphericities of the KA mixture. One can see that both volumes and
sphericities of this system demonstrate two peaks, i.e., the local
structures of the Voronoi cells have both different volume and
shape. It makes the KA mixture a much better glass-former than the
RSS. This result is in agreement with the results of the
simulations. Moreover, it proves that a binary mixture is a much
better glass-former than a monatomic system, such as an RSS.

In order to compare the results for different systems we make a
plot of the probability distribution of sphericity in scaled
coordinates: the value of sphericity is scaled to the value of an
ideal crystal (1.167 for FCC and 1.48 for the diamond structure)
and the distribution itself is normalized to make the height of
the peak equal to unity. We compare the width at half-height
$\Gamma$ of the distributions for the different systems. In the
case of the RSS the peak is split into two subpeaks above the
half-height. We scale the height of the peak on the value of the
higher subpeak. In the case of the KA mixture the splitting is
more pronounced and the smaller peak is below the half-height.
Because of this we use the half-height of the main peak. However,
in both cases of the RSS and KA mixture the distribution cannot be
approximated by a single peak and values of $\Gamma$ cannot be
strictly compared with those of the polydisperse LJ and amorphous
silicon.

\begin{figure}
\includegraphics[width=8cm, height=8cm]{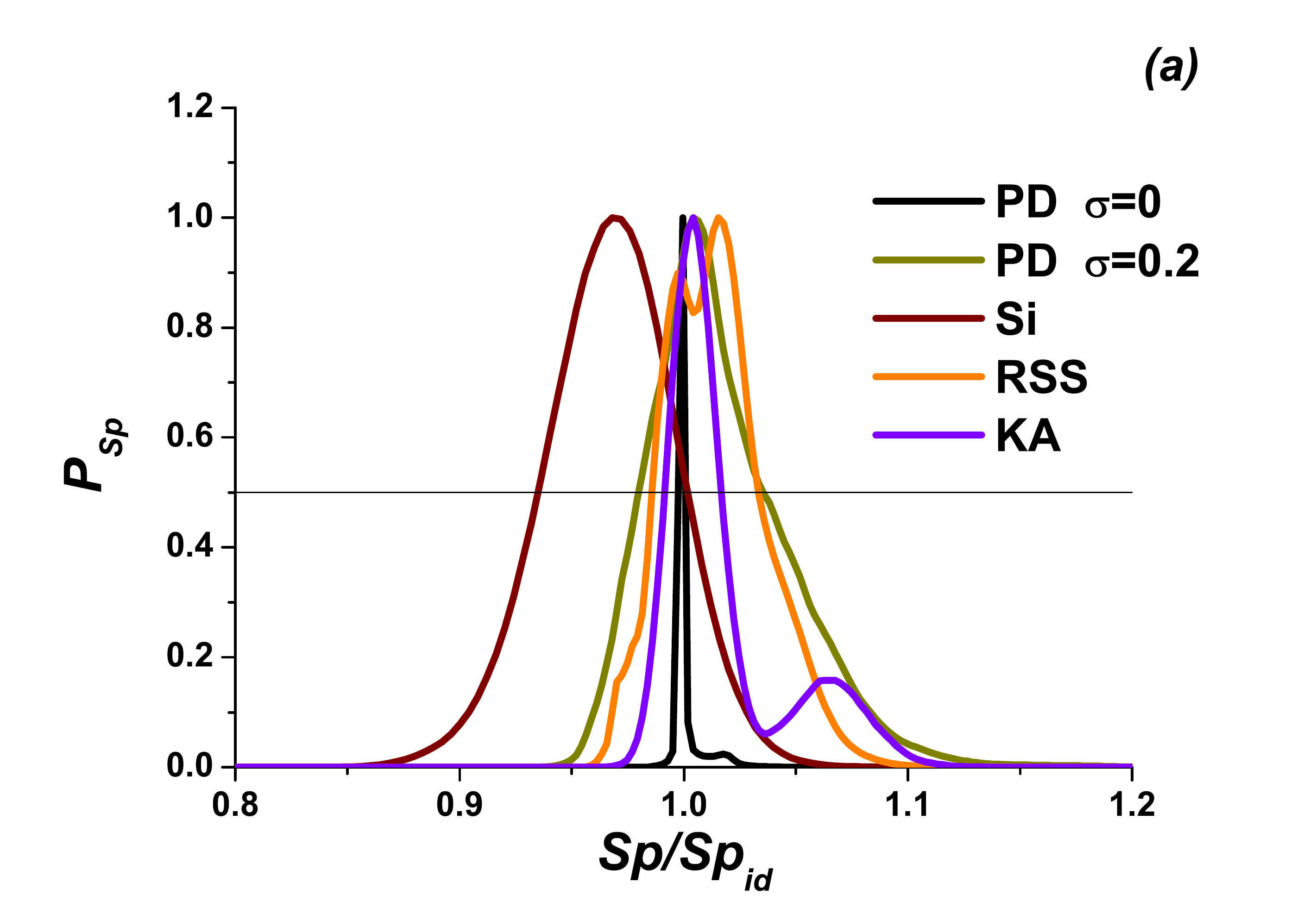}
\includegraphics[width=8cm, height=8cm]{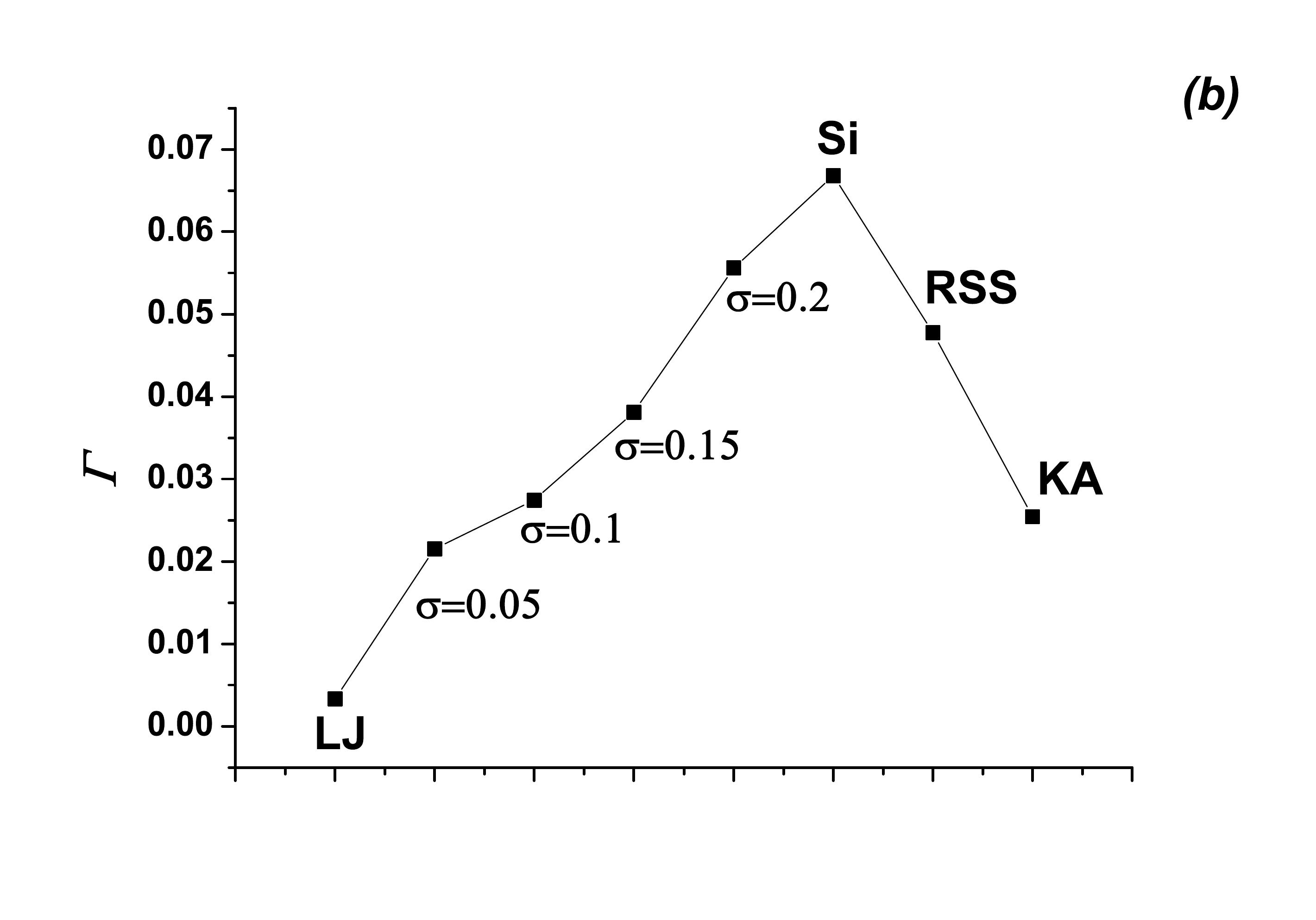}

\caption{\label{gamma} The probability distribution of
sphericities in scaled coordinates: $Sp/Sp{id}$, where
$Sp_{id}=1.48$ for amorphous silicon (the diamond structure) and
$Sp_{id}=1.167$ for all other systems (the FCC structure). (b) The
width at the half-height of the distribution of different systems
(see the text).}
\end{figure}

Figures \ref{gamma} (a) and (b) confirm the mechanism discussed
above. One can see that the width of probability distribution is
very small in the case of the pure LJ system, but if the system
becomes polydisperse, width $\Gamma$ rapidly increases. Width
$\Gamma$ is even higher for amorphous silicon, which means that
silicon has a strong tendency to amorphization. The width of the
probability distribution of the RSS and KA mixture is not a
well-defined quantity and we do not discuss it here.

The main arguments of the present study are based on the analysis
of the probability distribution of the volumes and sphericities of
the Voronoi polyhedra of different systems. Similar analysis of
the Voronoi cells of liquids has been performed in a large body of
studies, including the Lennard-Jones system
\cite{abascal1,lj-vor,voloshin}, water and similar systems
($H_2S$, $HF$) \cite{ruocco1,ruocco2,shih,jedl1,jedl2}, amorphous
silicon \cite{si-vor} and ionic melts \cite{abascal2,ruff}.
However, all these papers except Ref. \cite{si-vor} deal mostly
with the liquid phase and therefore the probability distributions
of the properties of Voronoi polyhedra demonstrated a single peak
only. Our study showed that the behavior of the probability
distribution of the different properties of the Voronoi cells
could be more complex in the case of glassy systems. Importantly,
the probability distribution should be accompanied by other
characteristics of the system to exclude the possibility of
partial crystallization of the system, which could lead to two
peak distributions. In the present work we control this by
calculating the RDFs, MSDs and ISFs.

\section{Conclusions}

In this paper we analyze the distribution of the volume and shape
of the Voronoi cells in several glass-forming systems. We show
that two types of the distributions are possible. Some of the
glass-formers demonstrate a single but very spread peak of the
sphericity parameter. It means that the Voronoi cells of the
particles have a wide distribution of shapes, which prevents the
crystallization of the system. In some other systems the
probability distribution of the sphericity of the Voronoi cells
splits into two peaks, i.e., two types of cells appear in the
system. From this observation we conclude that two mechanisms of
glass formation in monatomic systems are possible: via the
spreading of a single local structure and via the splitting of a
local structure into two types. This mechanism is also responsible
for glass transition in binary mixtures.

These results shed light on the microscopic nature of glass
transition in monatomic systems.

This work was carried out using computing resources of the federal
collective usage center "Complex for simulation and data
processing for mega-science facilities" at NRC "Kurchatov
Institute", http://ckp.nrcki.ru, and supercomputers at Joint
Supercomputer Center of the Russian Academy of Sciences (JSCC
RAS). The work was supported by the Russian Science
Foundation (Grants No 19-12-00092).

\section{A Systems and Methods}

\subsection{Polydisperse Lennard-Jones particles}

The Lennard-Jones potential is defined as

\begin{equation}
U(r)=4 \varepsilon \left( \left(\frac{\sigma_i}{r} \right)^{12} -
\left(\frac{\sigma_i}{r} \right)^6 \right),
\end{equation}
where $\sigma_i$ is the diameter of the $i-$th particle. Parameter
$\varepsilon$ serves as an energy scale.

The diameters of particles are distributed around $\sigma_i=1.0$
with the Gaussian distribution with dispersion $\sigma$. Several
values of $\sigma$ are used: $\sigma=0.0$, $0.05$, $0.1$, $0.15$
and $0.2$. The simulation setup for all values of polydispersity
is the same.

Systems of $32000$ particles in a cubic box with periodic boundary
conditions (PBC) are used. The time step is set to $dt=0.001$. The
initial structure is obtained as a high temperature configuration
of liquid. Then we equilibrate the systems at temperatures from
$T_{min}=0.1$ to $T_{max}=1.5$ with a step in temperature $dT=0.1$
for $5 \cdot 10^7$ steps in the canonical ensemble (constant
number of particles N, volume V and temperature T). After that, we
perform simulations for more than $5 \cdot 10^7$ steps in a micro
canonical ensemble (constant number of particles N, volume V and
internal energy E). At this stage we calculate the properties of
the system: equation of state (EOS), radial distribution functions
(RDF), mean square displacement (MSD) and intermediate scattering
functions (ISF). The wave vector for the ISF is chosen as the
first maximum of the static structure factor.

\subsection{Amorphous silicon}

In the case of amorphous silicon, a system of 8000 particles is
used. The initial configuration is obtained by melting the diamond
structure of the system at $T=8000$ K and $P=1.0$ bar. Then we
simulate the system at constant pressure $P=1.0$ bar and
temperatures from $300$ to $1500$ K with step $100$ K for $1 \cdot
10^7$ steps with time step 1fs.

Additional simulations were performed for temperatures from
$T=2000$ K to $4000$ K with step $dT=500$ K. It was found that at
$T=2000$ K the system spontaneously crystallized. A video of this
crystallization is given here. Experimentally the melting point of
silicon is 1687 K, which is much below the crystallization point
of our simulation. Because of this we think that the model we
employ is not valid for high temperature. However, as was shown in
Ref. \cite{vink} this potential properly described the properties
of amorphous silicon. Because of this only the data for the low
temperature amorphous phase are discussed in the present paper.

In the isobaric-isothermal simulation we determine the equilibrium
density at given temperature. Then we perform simulations in the
canonical ensemble for another $1 \cdot 10^7$ steps for
equilibration and more $1 \cdot 10^8$ steps in the NVE ensemble
for calculation of the properties of the system.

\subsection{The Repulsive Shoulder System}

The Repulsive Shoulder System (RSS) is defined by the potential

\begin{equation}\label{pot}
U(r)/ \varepsilon = \left( \frac{\sigma}{r}
\right)^{14}+0.5\left(1- tanh(k(r-\sigma_1)) \right),
\end{equation}
where parameters $\varepsilon$ and $\sigma$ set energy and length
scales, $k=10.0$ and parameter $\sigma_1=1.35$ sets the width of
the repulsive shoulder.

The behavior of this system was carefully studied in a number of
papers (see Refs. \cite{s1,s2,s3,s4} and references therein). It
was shown that this system demonstrated a very complex phase
diagram which strongly depended on parameter $\sigma_1$ and
numerous water-like anomalies. In particular, in Ref. \cite{we1}
it was found that at $\sigma_1=1.35$ the system experienced glass
transition. Later on, this result was confirmed in Ref.
\cite{rysch}.

In the present study we simulate a system of 32000 particles in a
cubic box with PBC. The initial configuration is high temperature
liquid. Then it is infinitely fast quenched to the desired
temperature. Firstly, the system is equilibrated in the NVT
ensemble for $2 \cdot 10^7$ steps. After that the system is
simulated for another $5 cdot 10^7$ steps in the NVE ensemble for
calculation of its properties. The temperatures are from $0.06$ to
$0.2$ with step $dT=0.01$.

\subsection{The Kob-Andersen mixture}

The Kob-Andersen (KA) mixture is a system of LJ particles of two
types, A and B \cite{ka1}. The interaction parameters are
$\varepsilon_{AA}=1.0$, $\sigma_{AA}=1.0$, $\varepsilon_{AB}=1.5$,
$\sigma_{AB}=0.8$, $\varepsilon_{BB}=0.5$, $sigma_{AA}=0.88$. The
concentration of A particles is $80 \%$.

We simulate the KA mixture at number density $\rho=N/V=1.2$, where
$N=N_A+N_B$ is the total number of particles in the system. The
simulation setup is very similar to that of the polydisperse LJ
system. The initial configuration is obtained as high temperature
liquid. Then the system is quenched immediately to temperatures
$T=0.1$, $0.2-1.2$ with step $dT=0.2$. The system is equilibrated
in the NVT ensemble for $2 \cdot 10^7$ steps. More $2 \cdot 10^7$
steps are performed in the NVE ensemble for calculation of
properties.

\subsection{Units}

In the case of amorphous silicon, we use physical units of
measurements, i.e., temperature is measured in Kelvins, pressure
in bars, distance in Angstroms, etc.

In all other cases we use reduced units related to the interaction
potential, i.e., parameters $\varepsilon$ and $\sigma$ are used as
units of energy and length. All other quantities are expressed in
units based on these scales.

\section{B Results and Discussion}

In this section we show the radial distribution functions (RDFs),
mean square displacement (MSD) and intermediate scattering
functions (ISFs) of the systems under consideration.

\subsection{The Polydisperse Lennard-Jones system}

In this part of the paper, we show the RDFs, MSDs and ISFs of the
polydisperse LJ systems with different degrees of polydispersity:
$\sigma=0$ (the monodisperse system), $\sigma=0.05$, $0.1$, $0.15$
and $0.2$.

\begin{figure}
\includegraphics[width=8cm, height=8cm]{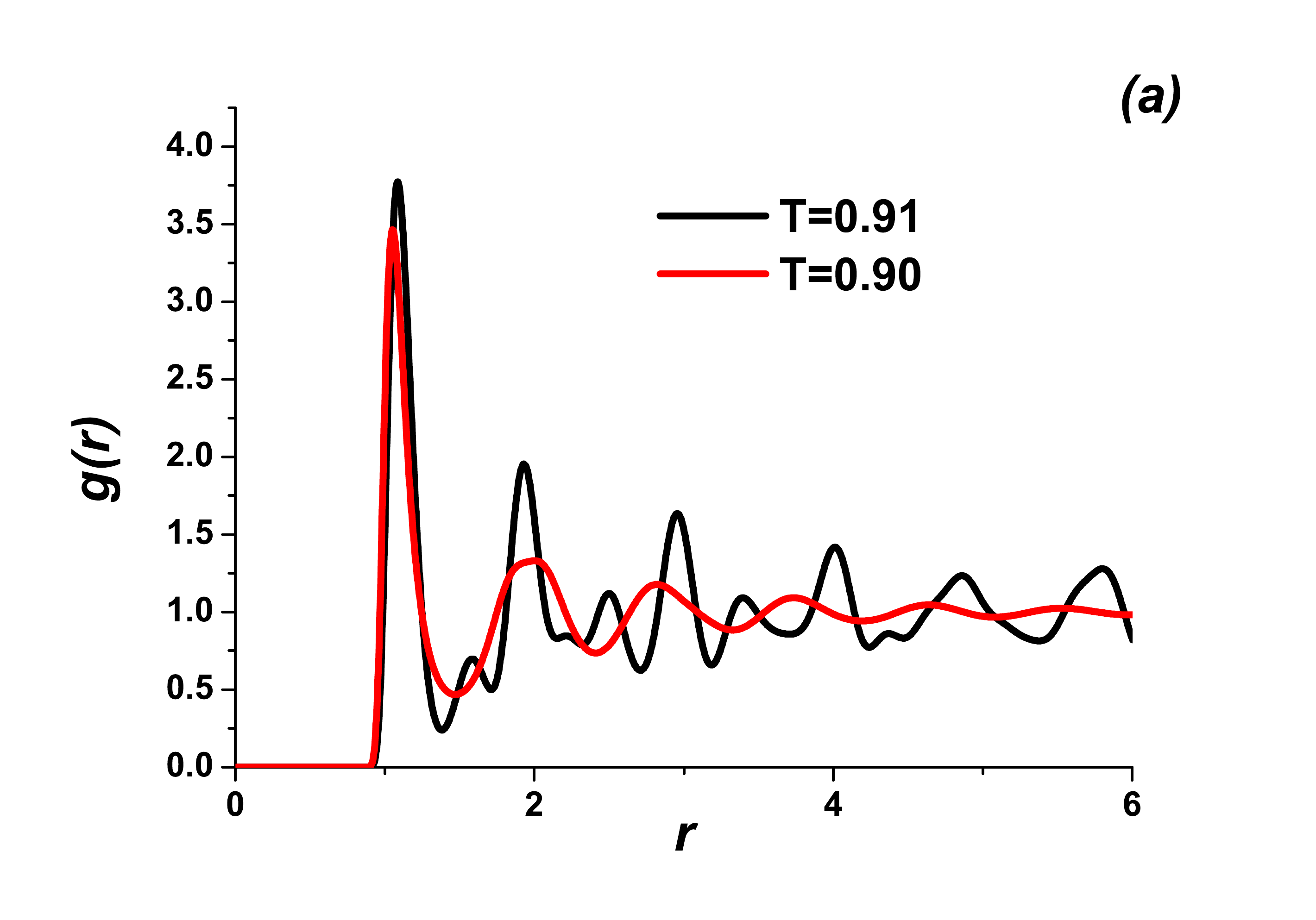}

\includegraphics[width=8cm, height=8cm]{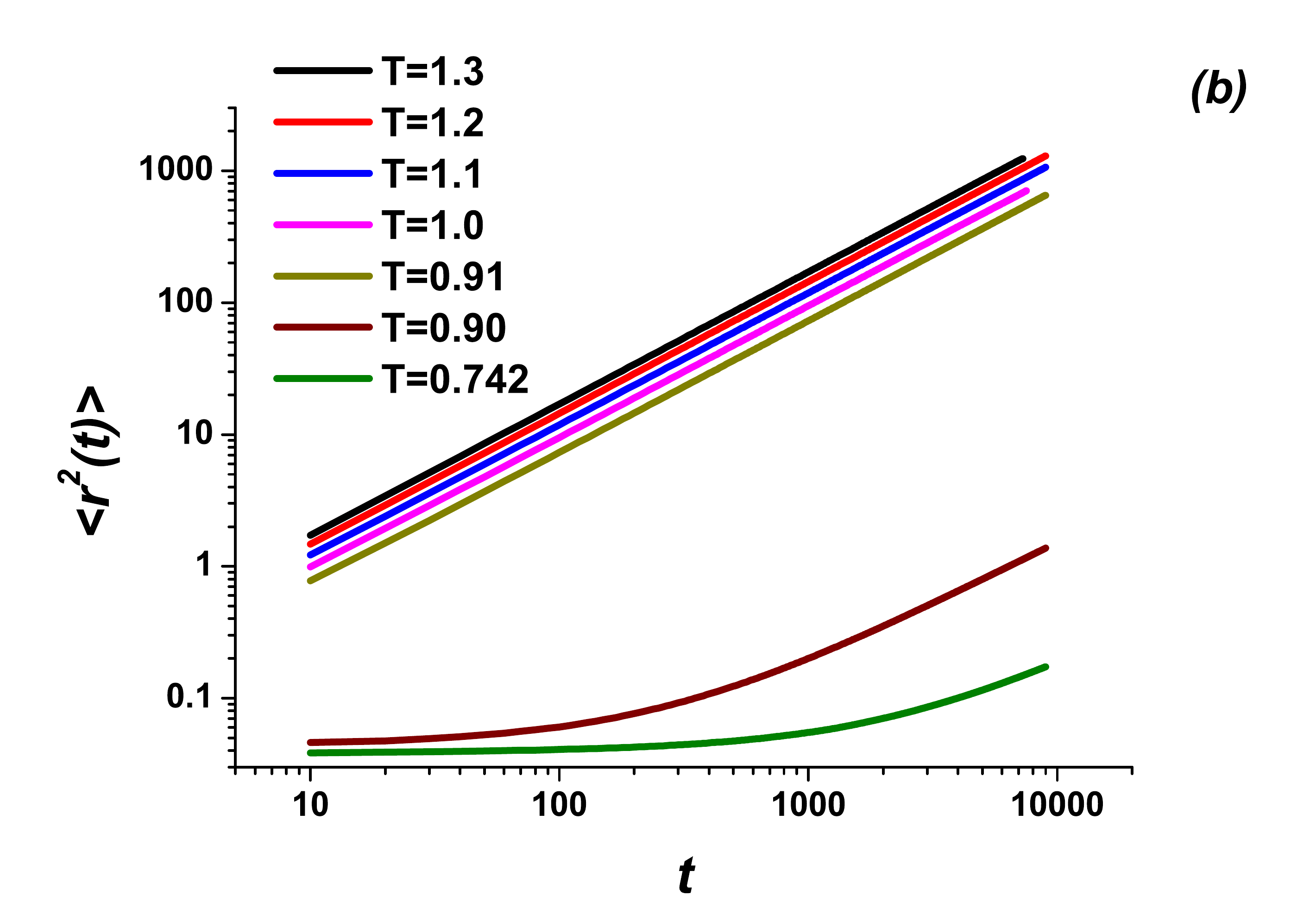}

\includegraphics[width=8cm, height=8cm]{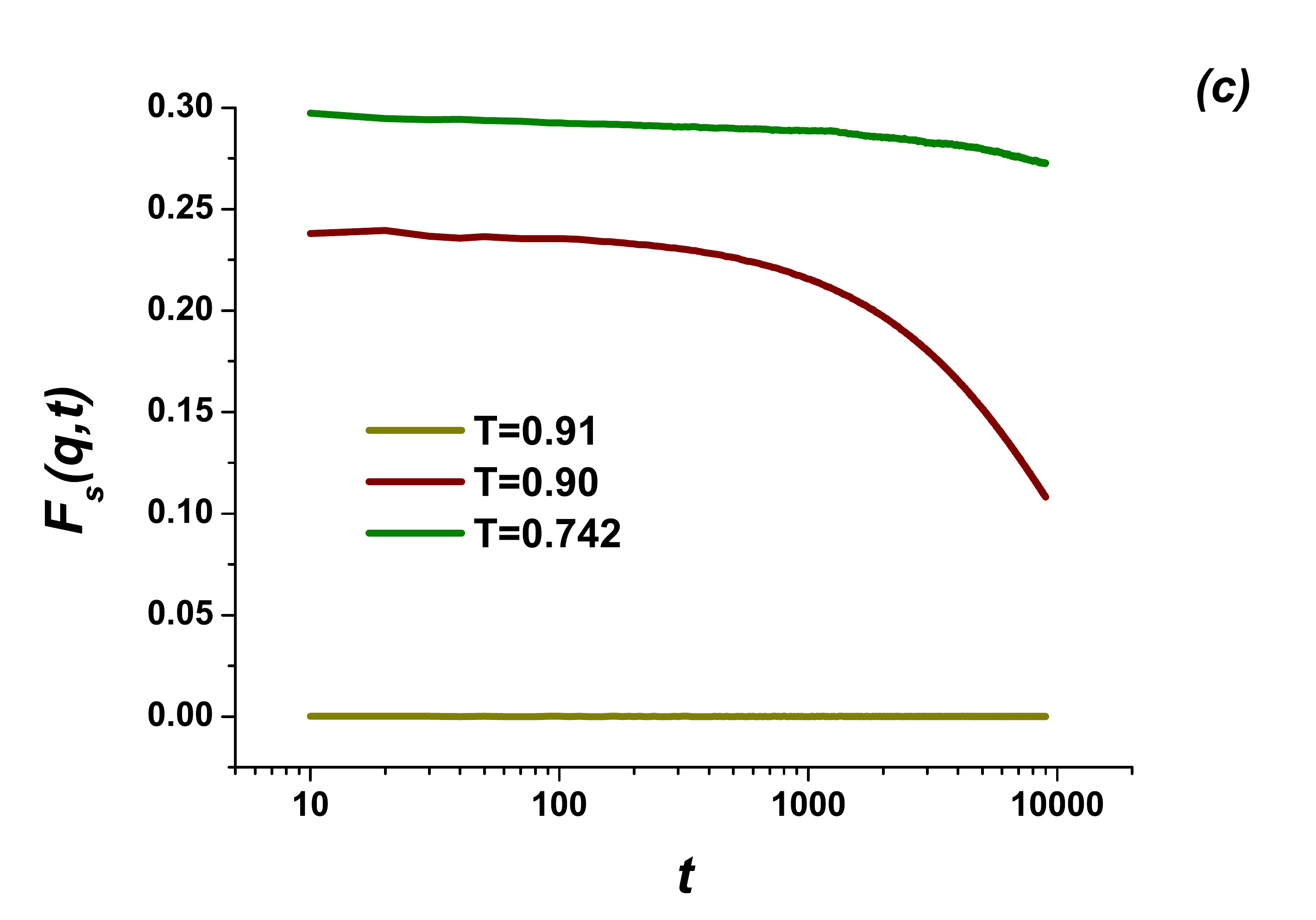}

\caption{\label{pd0-app} (a) The radial distribution functions,
(b) mean square displacement and (c) intermediate scattering
functions of the monodisperse LJ system at $\rho=1.0$ and a set of
temperatures. The wave vector for the ISFs is $k=1.58$.}
\end{figure}

In the case of the monodisperse system the crystallization is
easily visible from the behavior of the RDFs (Fig.
\ref{pd0-app}(a)). The MSD and ISFs of this system are given in
Fig. \ref{pd0-app} (b) and (c).

\begin{figure}
\includegraphics[width=8cm, height=8cm]{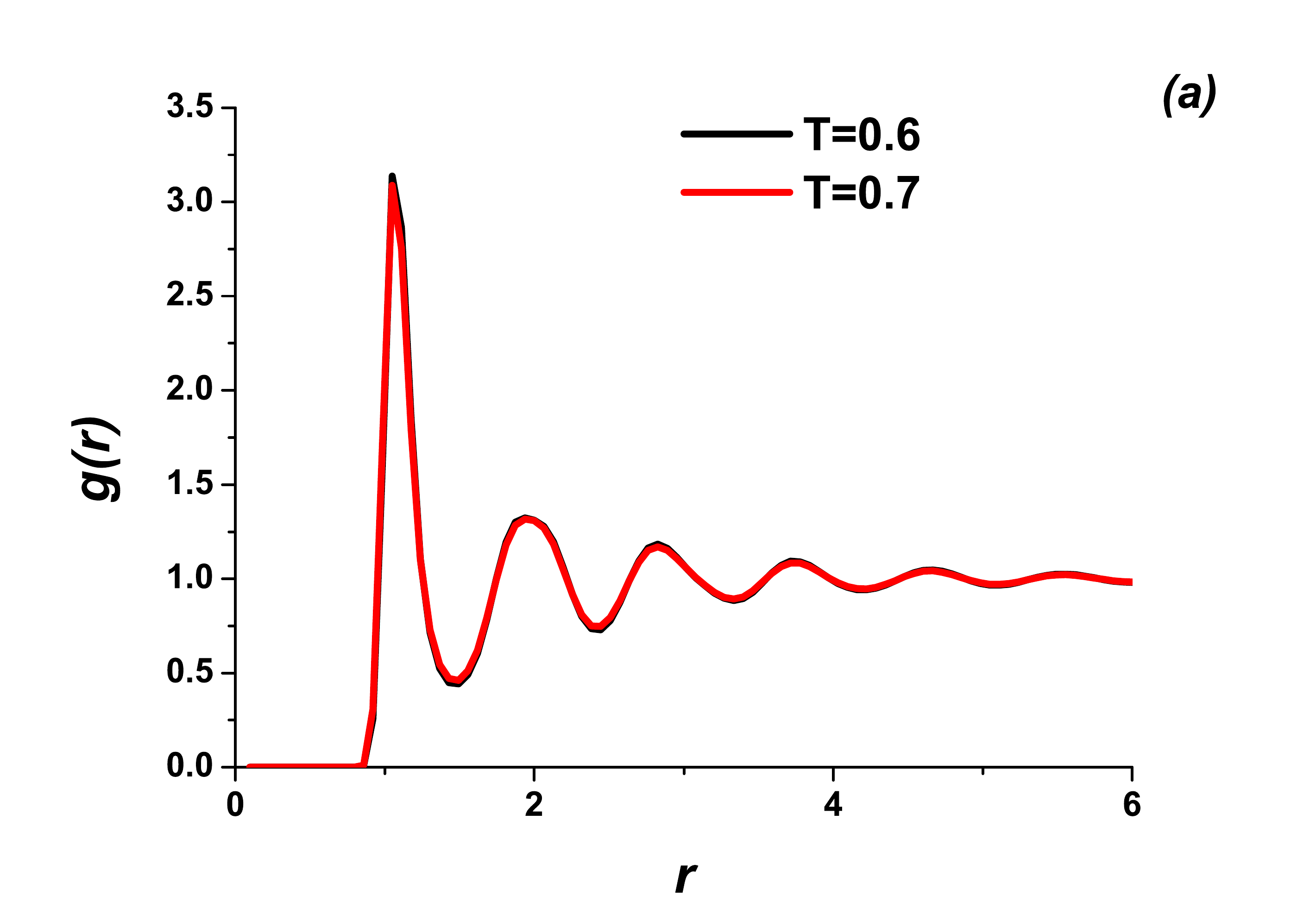}

\includegraphics[width=8cm, height=8cm]{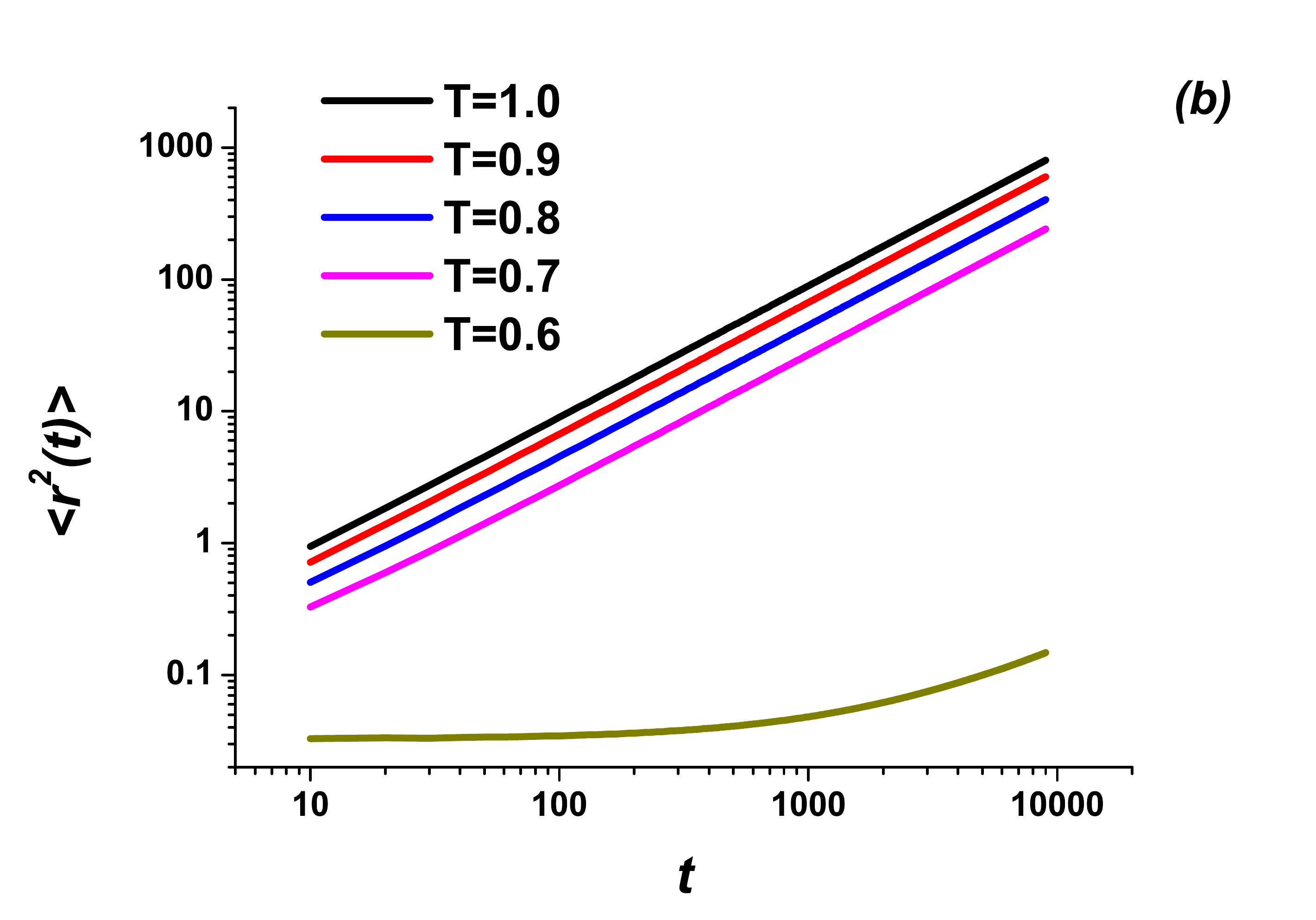}

\includegraphics[width=8cm, height=8cm]{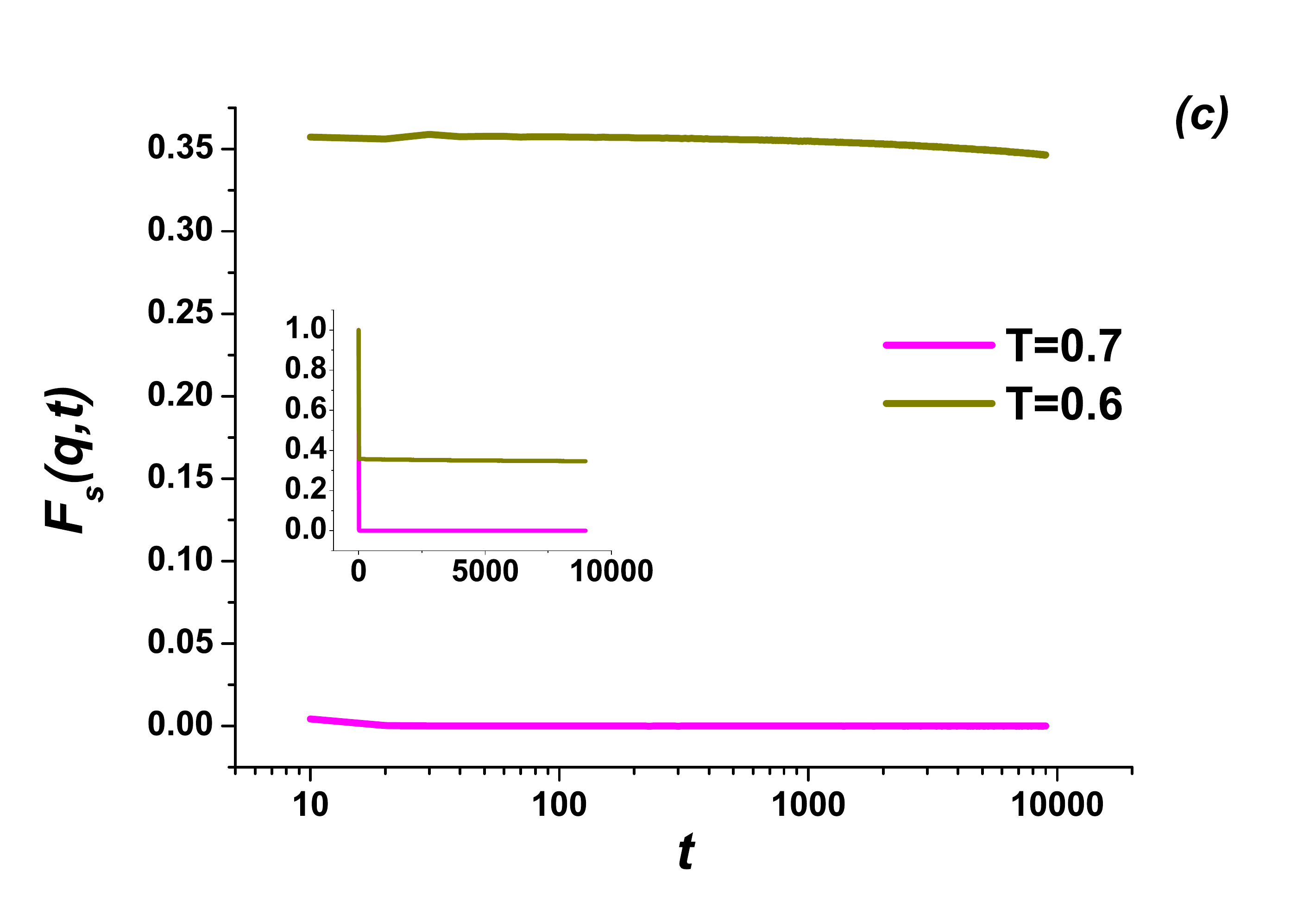}

\caption{\label{pd005-app} (a) The radial distribution functions,
(b) mean square displacement and (c) intermediate scattering
functions of the polydisperse LJ system at $\rho=1.0$ and a set of
temperatures. The degree of polydispersity is $5 \%$. The wave
vector for the ISFs is $k=0.9$ $\AA$.}
\end{figure}

From the EOS of the system with $5 \%$ polydispersity, shown in
the main text, we conclude that crystallization takes place at
temperature about $T=0.6$. However, the RDFs of the system remain
rather liquid-like even at the lowest temperatures (Fig.
\ref{pd005-app} (a)). However, crystallization of the system is
confirmed by the time correlation functions (TCFs) (Fig.
\ref{pd005-app} (b) and (c)). As shown in our recent work
\cite{qcgl}, in the case of crystallization the TCFs experience a
jump, while in glass transition they change continuously with
temperature. In the present case we do observe a jump. The
smearing of the RDFs is related to the formation of numerous small
crystallites which do not give a clear crystalline view of the
RDFs.

\begin{figure}
\includegraphics[width=8cm, height=8cm]{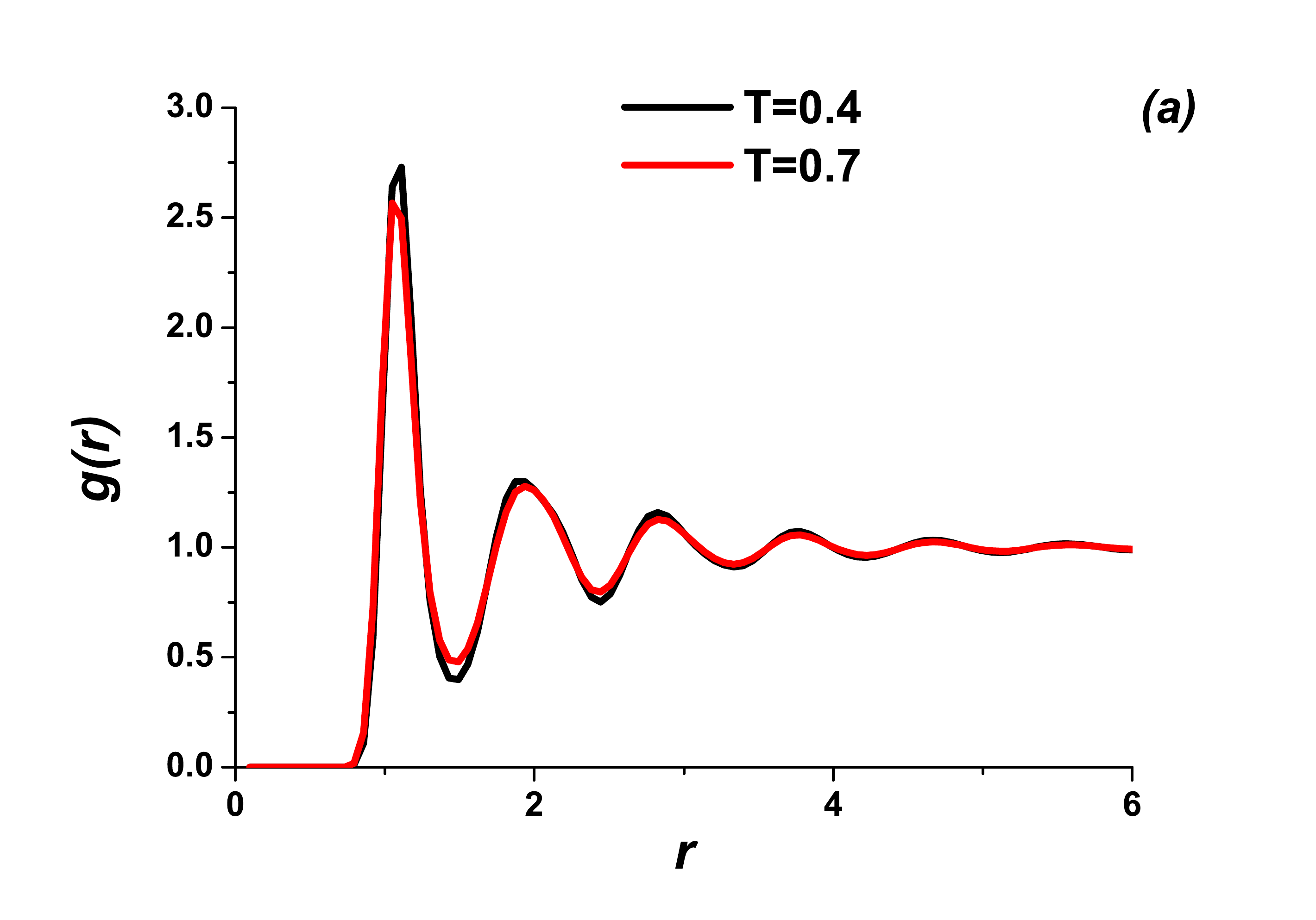}

\includegraphics[width=8cm, height=8cm]{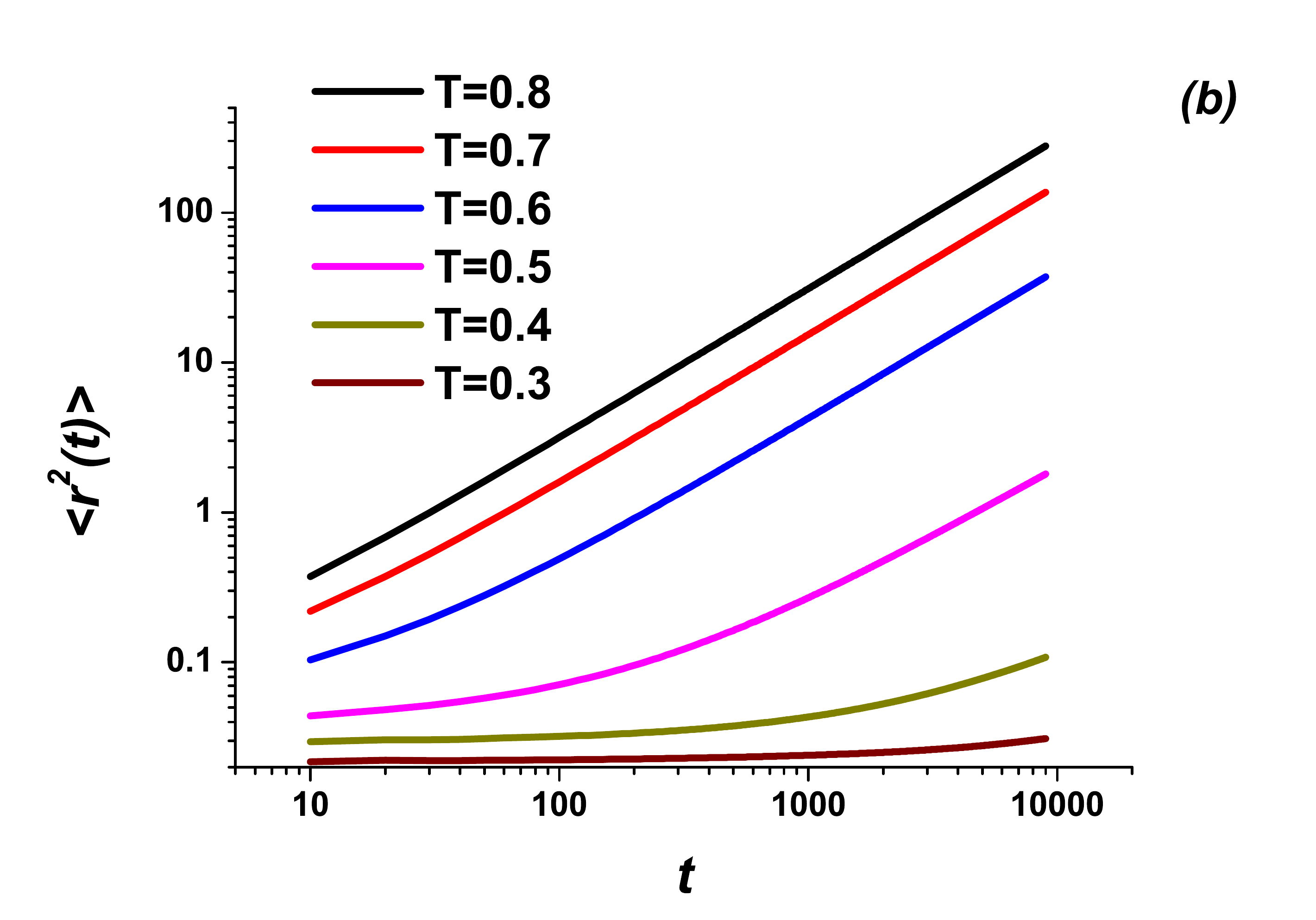}

\includegraphics[width=8cm, height=8cm]{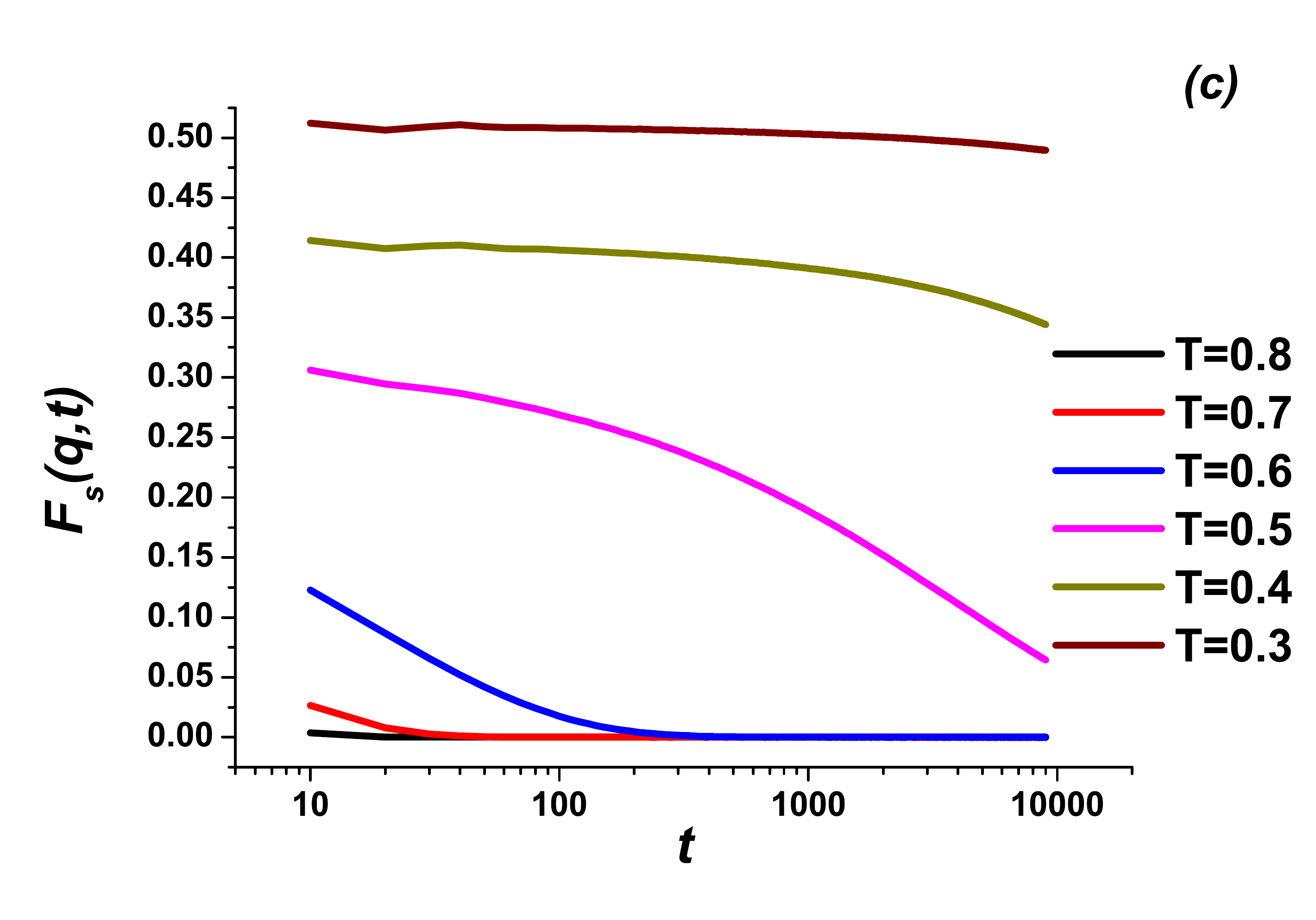}

\caption{\label{pd01-app} (a) The radial distribution functions,
(b) mean square displacement and (c) intermediate scattering
functions of the polydisperse LJ system at $\rho=1.0$ and a set of
temperatures. The degree of polydispersity is $10 \%$ The wave
vector for the ISF is $k=1.58$.}
\end{figure}

\begin{figure}
\includegraphics[width=8cm, height=8cm]{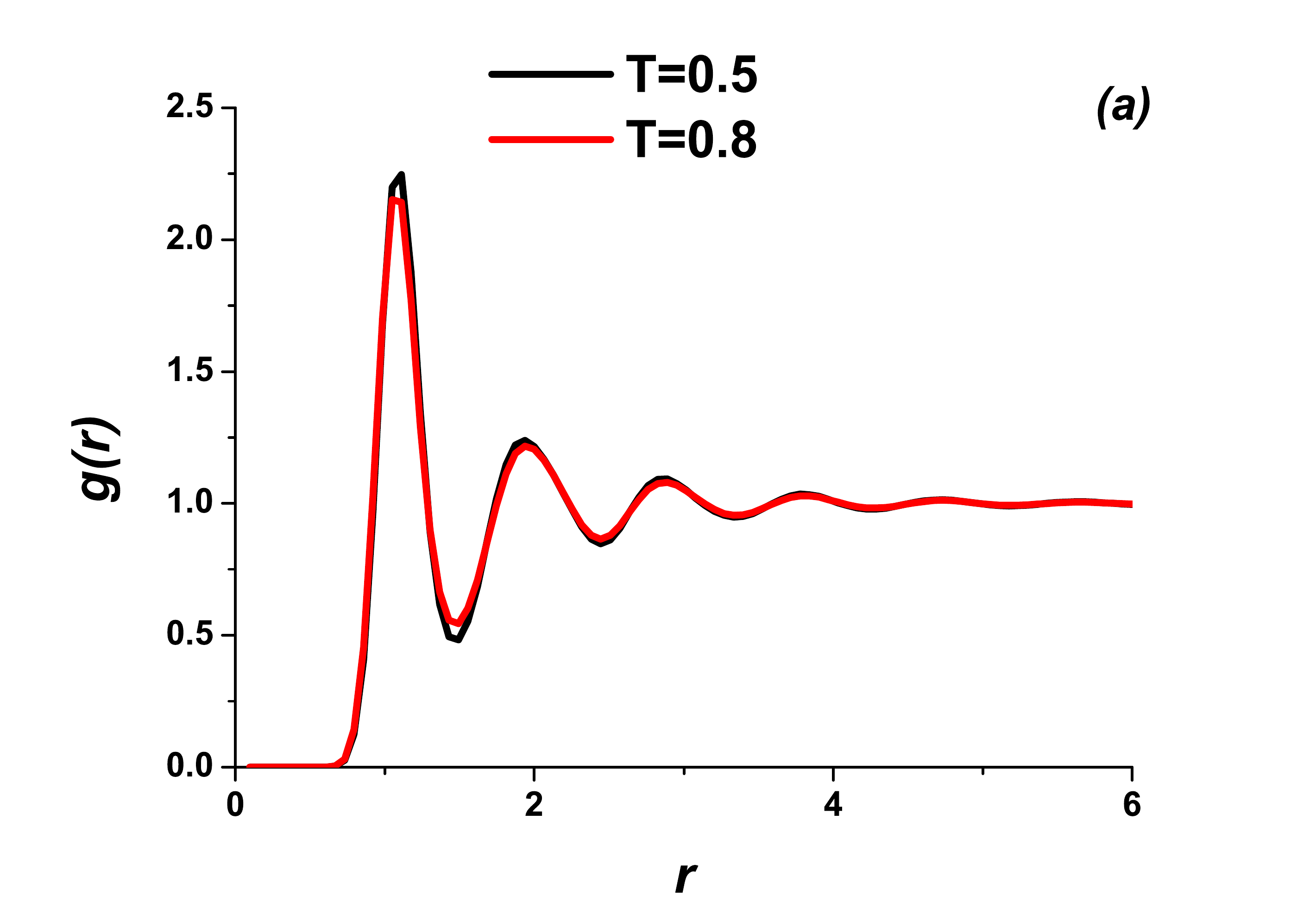}

\includegraphics[width=8cm, height=8cm]{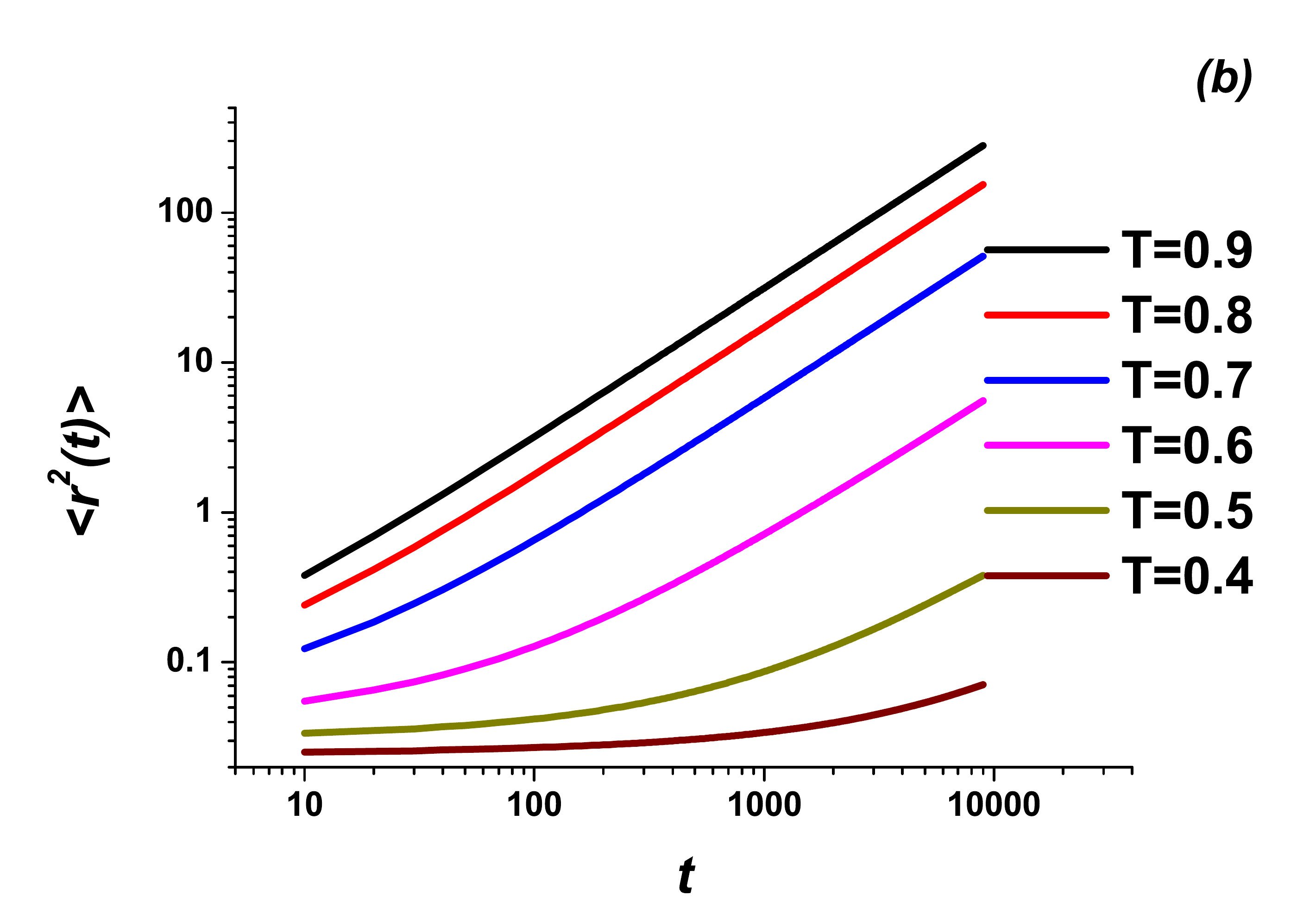}

\includegraphics[width=8cm, height=8cm]{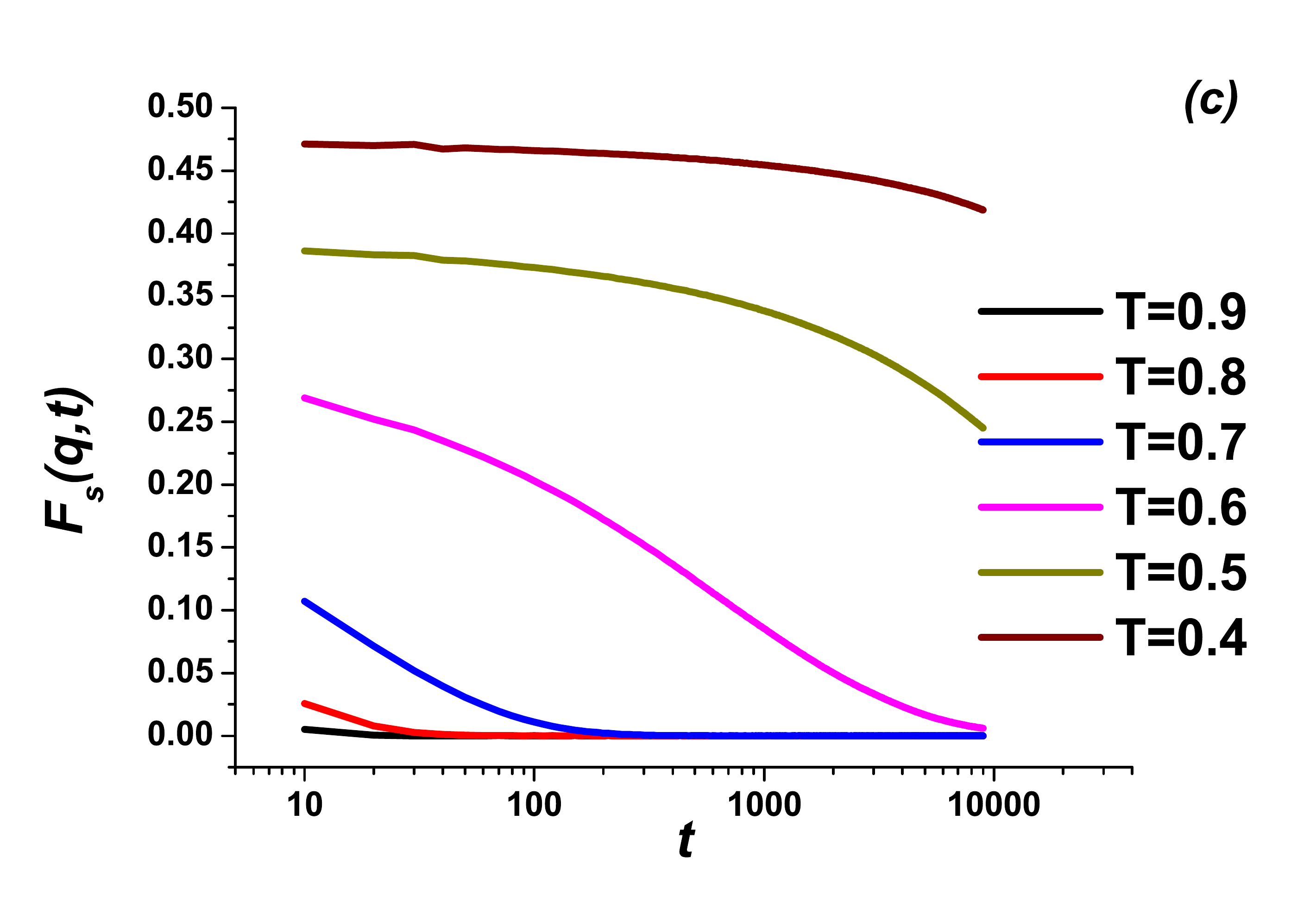}

\caption{\label{pd015-app} (a) The radial distribution functions,
(b) mean square displacement and (c) intermediate scattering
functions of the polydisperse LJ system at $\rho=1.0$ and a set of
temperatures. The degree of polydispersity is $15 \%$ The wave
vector for the ISFs is $k=1.58$.}
\end{figure}

\begin{figure}
\includegraphics[width=8cm, height=8cm]{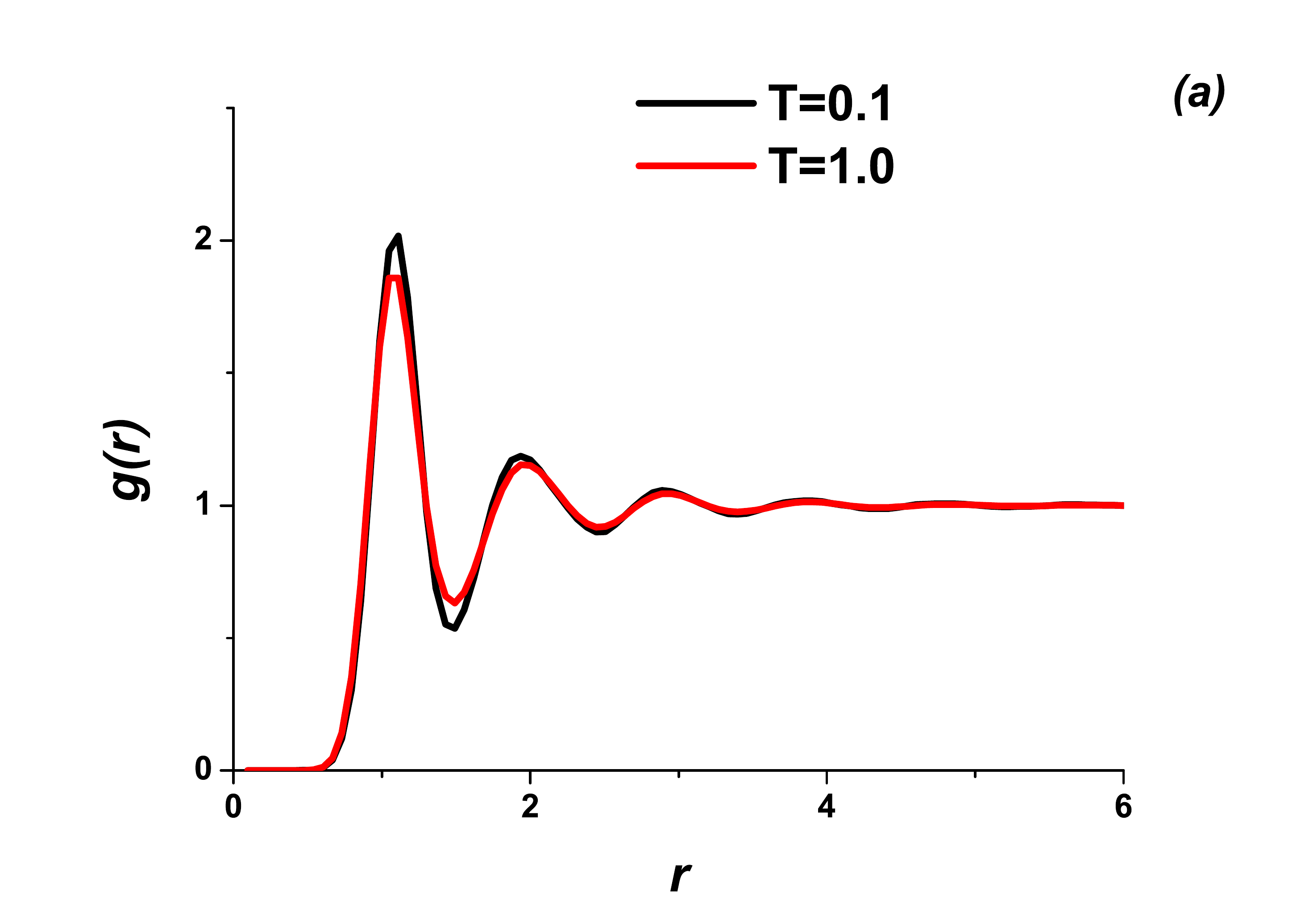}

\includegraphics[width=8cm, height=8cm]{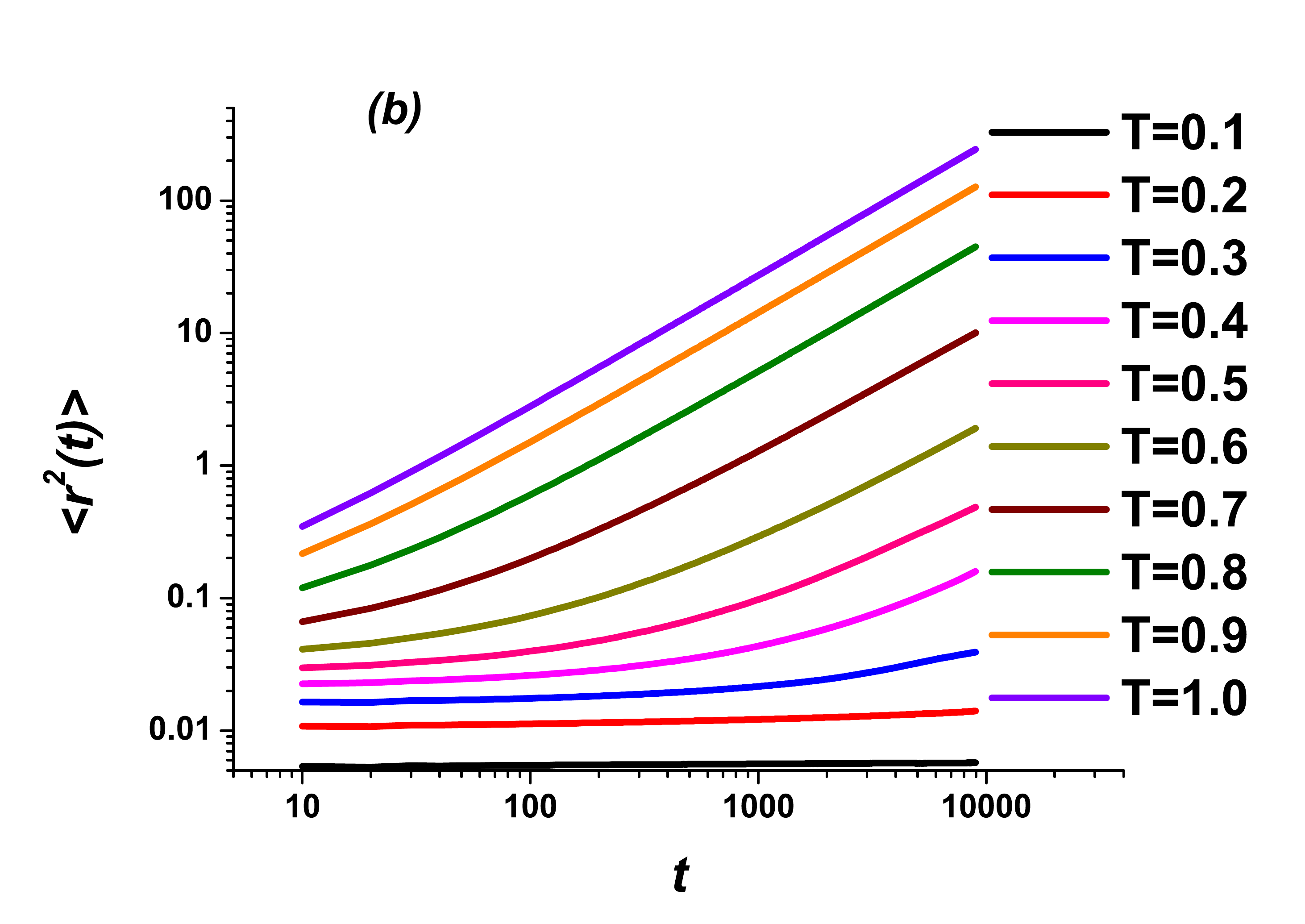}

\includegraphics[width=8cm, height=8cm]{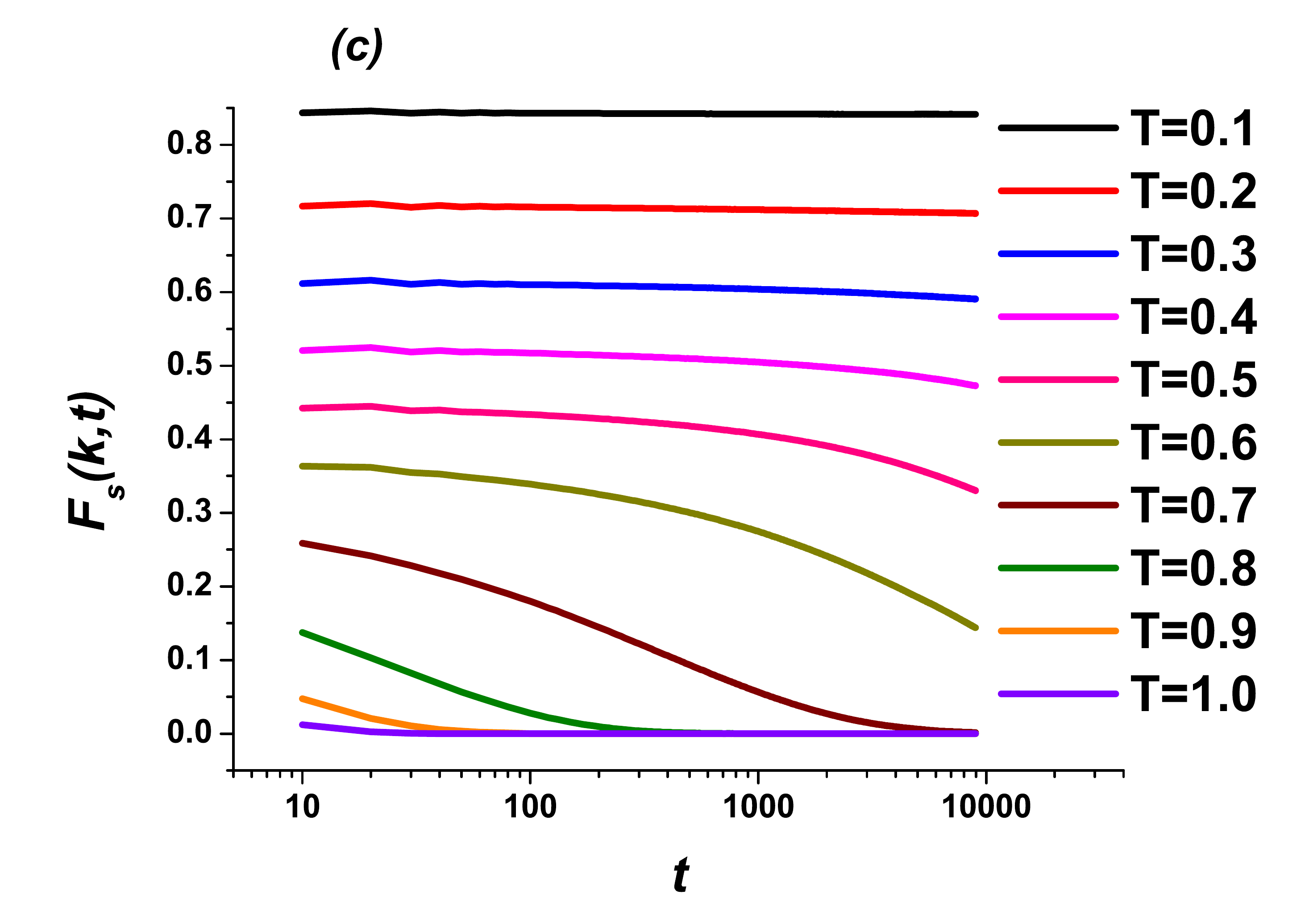}

\caption{\label{pd02-app} (a) The radial distribution functions,
(b) mean square displacement and (c) intermediate scattering
functions of the polydisperse LJ system at $\rho=1.0$ and a set of
temperatures. The degree of polydispersity is $20 \%$. The wave
vector for the ISFs is $k=1.58$.}
\end{figure}

In the case of polydispersity $10 \%$ and more both RDFs and TCFs
demonstrate glass transition, which can be seen in Figs.
\ref{pd01-app}, \ref{pd015-app} and \ref{pd02-app}.

\subsection{Amorphous silicon}

\begin{figure}
\includegraphics[width=8cm, height=8cm]{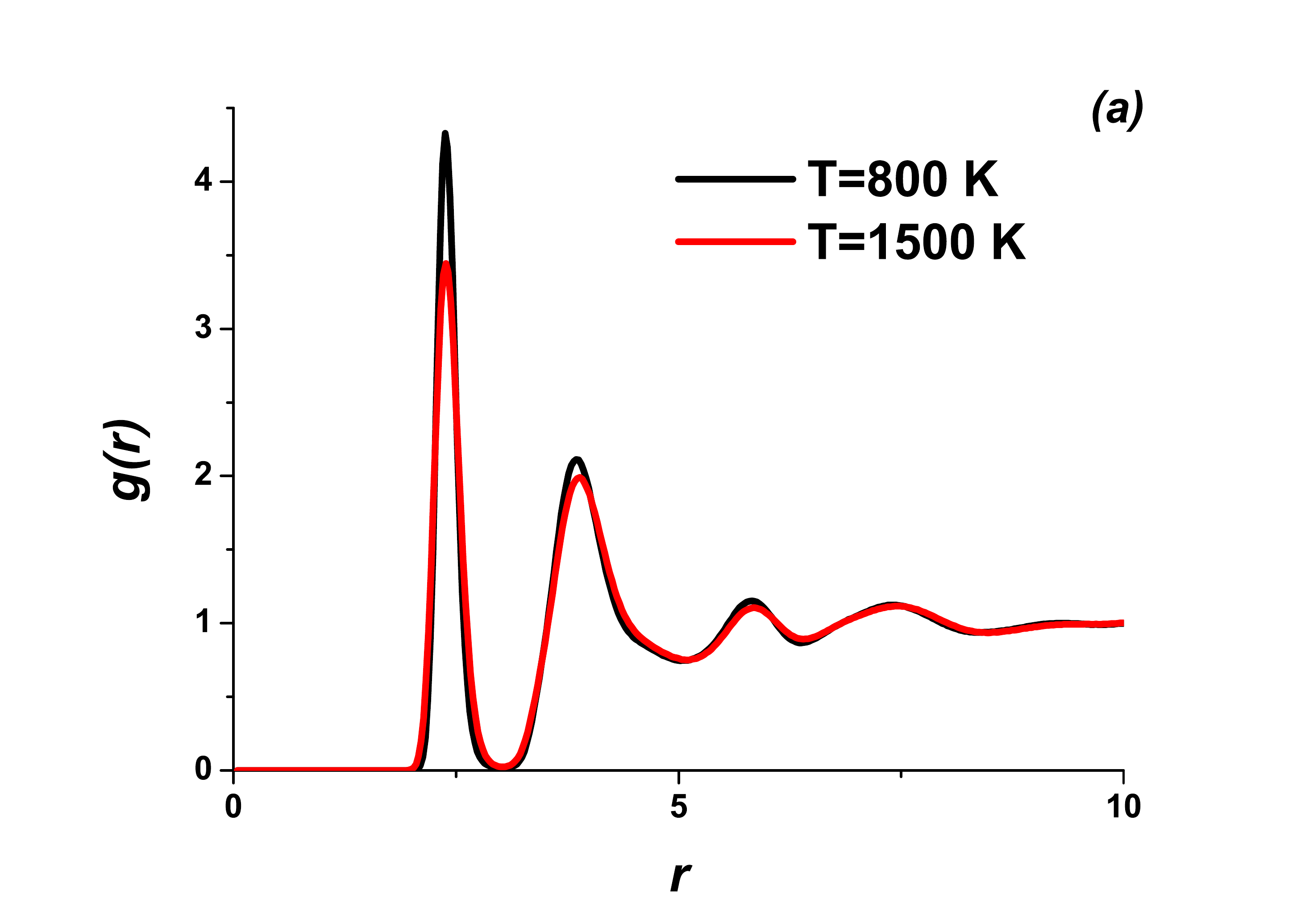}

\includegraphics[width=8cm, height=8cm]{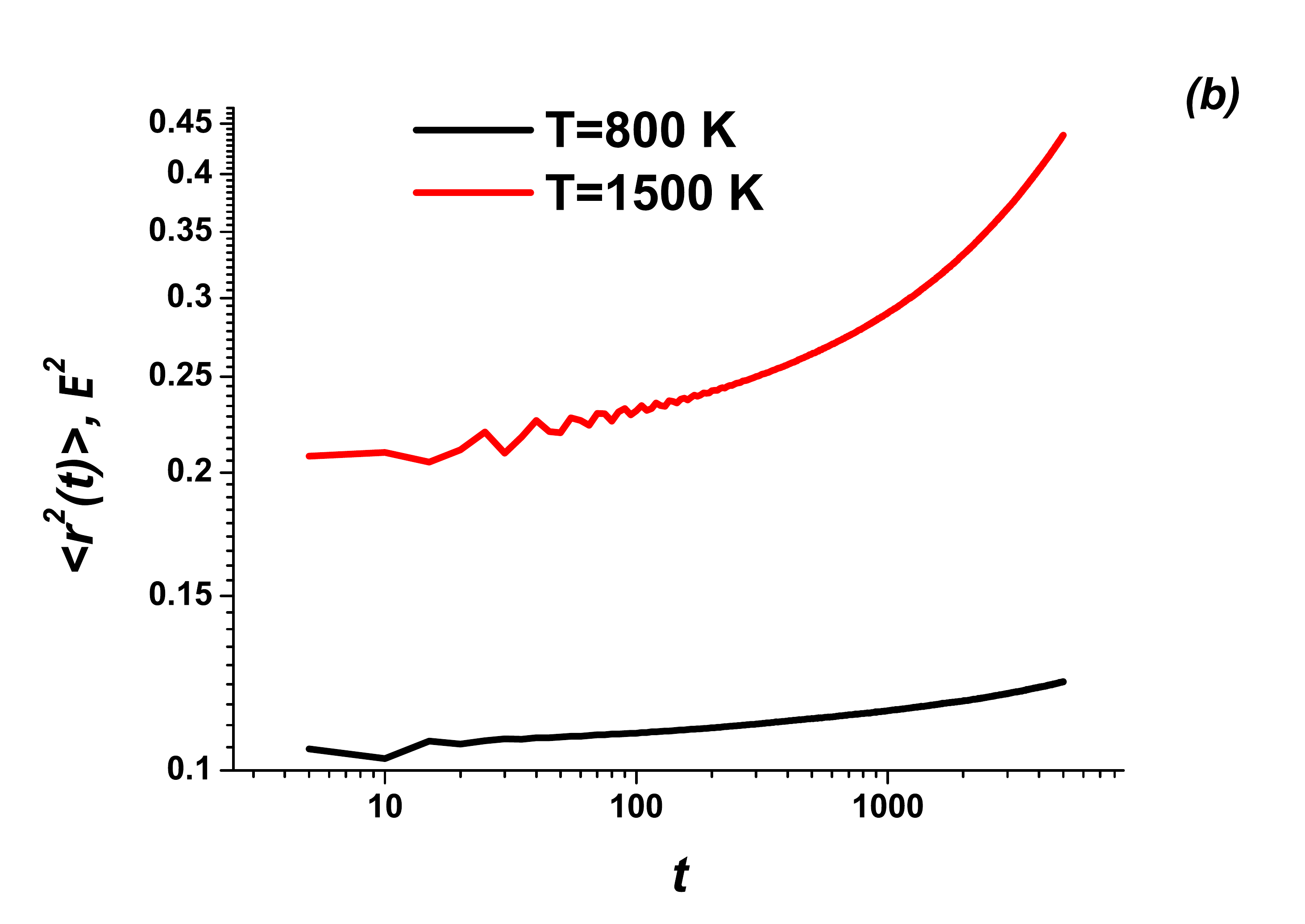}

\includegraphics[width=8cm, height=8cm]{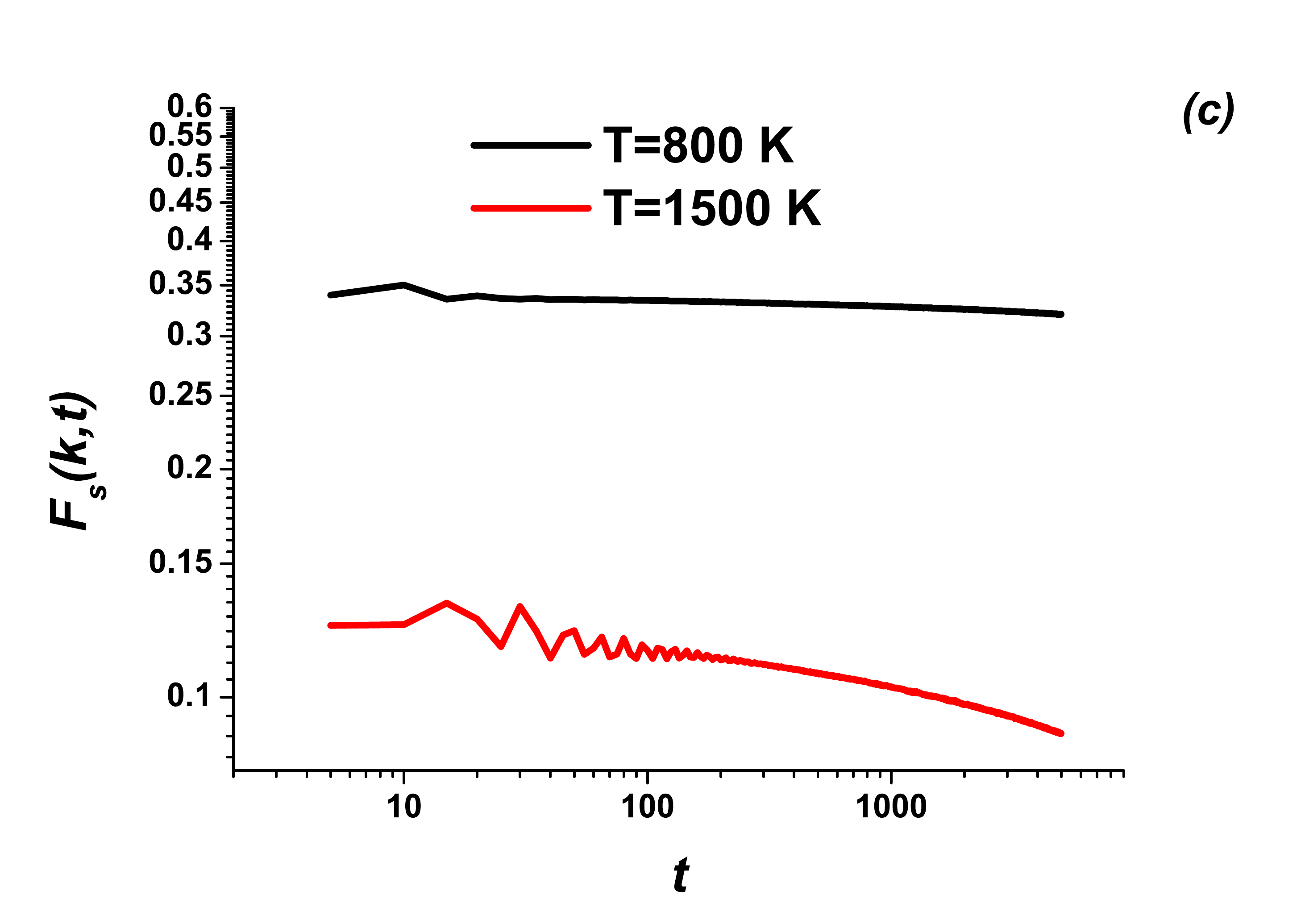}

\caption{\label{eos-pd-app} (a) The radial distribution functions,
(b) mean square displacement and (c) intermediate scattering
functions of amorphous silicon at two temperatures and ambient
pressure. The wave vector for the ISFs is $k=8.8$.}
\end{figure}

The radial distribution functions, mean square displacement and
intermediate scattering functions of amorphous silicon at two
temperatures and ambient pressure are given in Fig.
\ref{eos-pd-app} (a)-(c).

\subsection{The Repulsive Shoulder System}

Glass transition in the RSS is first reported in Ref. \cite{we1}.
Later on, it was confirmed in Ref. \cite{rysch}. The glass
transition temperature in \cite{we1} is found to be $T_g=0.07$.
The more elaborate calculations of Ref. \cite{rysch} report the
glass transition temperature to be $T_g=0.064$.

\begin{figure}
\includegraphics[width=8cm, height=8cm]{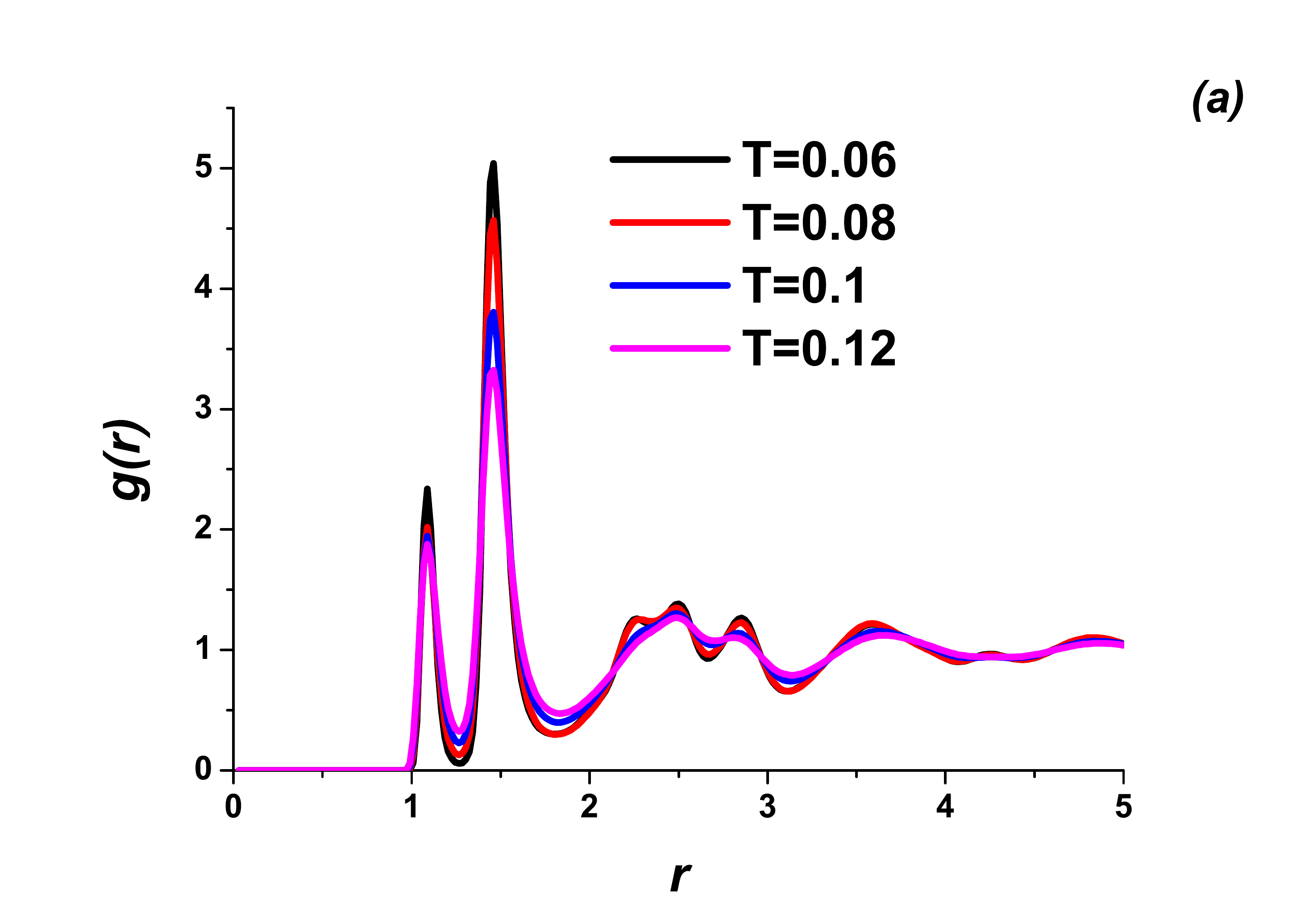}

\includegraphics[width=8cm, height=8cm]{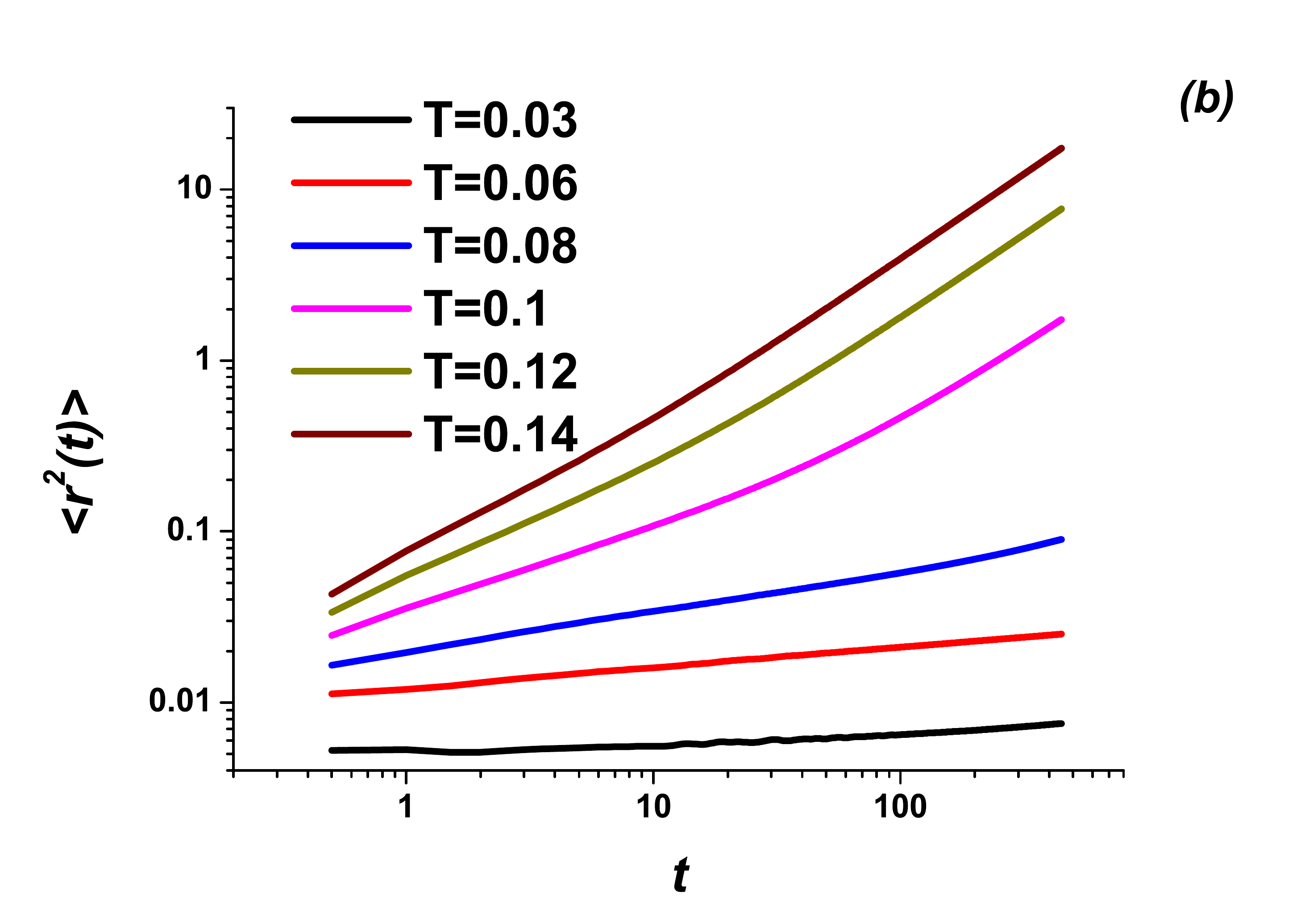}

\includegraphics[width=8cm, height=8cm]{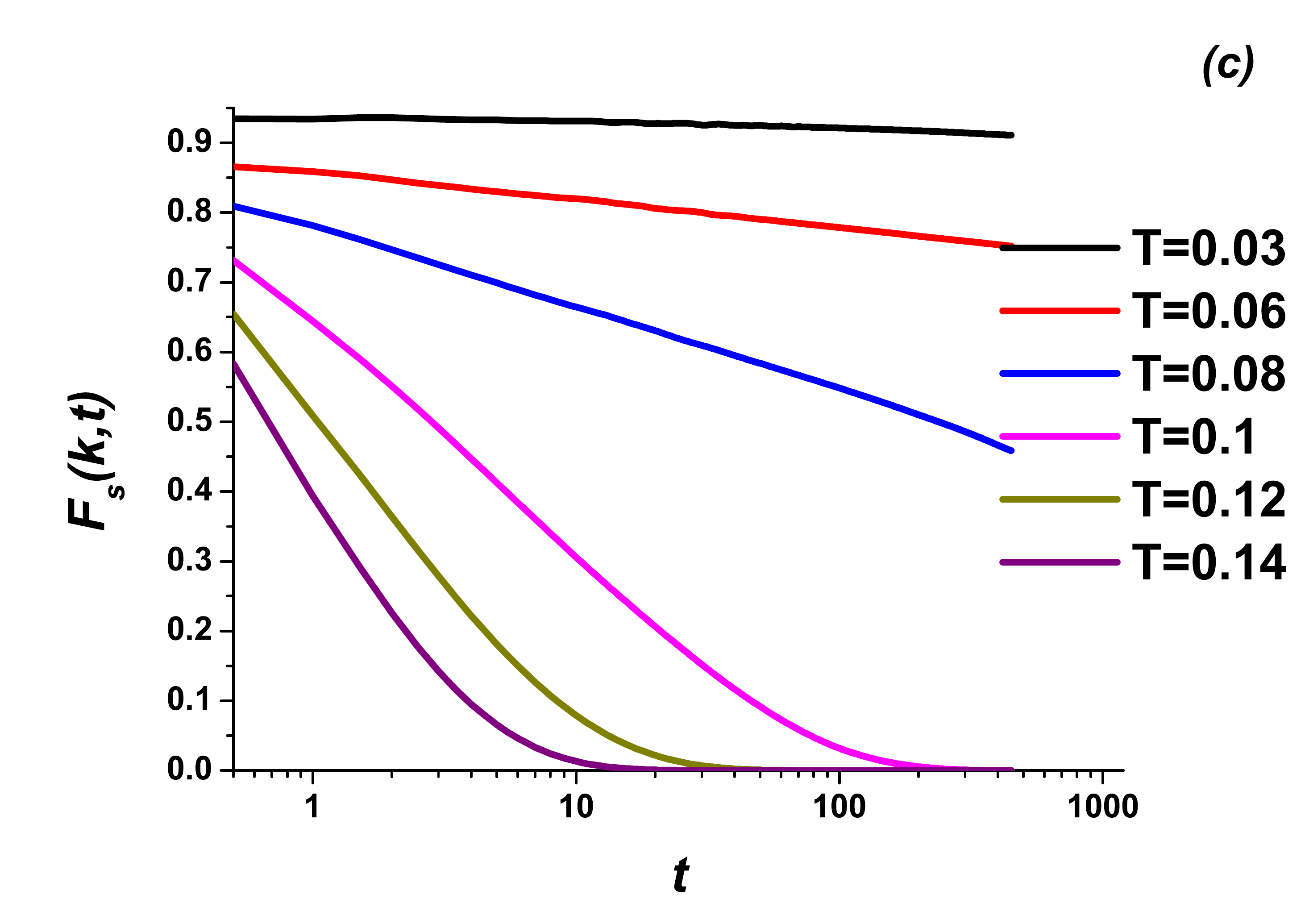}

\caption{\label{s135-app} (a) The radial distribution functions,
(b) mean square displacement and (c) intermediate scattering
functions of the RSS at $\rho=0.53$ and a set of temperatures. The
wave vector for the ISFs is $k=8.8$.}
\end{figure}

The lowest temperature of the calculations of this work is
$T_{min}=0.06$, i.e., below glass transition. The RDFs, MSDs and
ISFs of this system at $\rho=0.53$ and a set of temperatures are
given in Fig. \ref{s135-app} (a)-(c).

\subsection{The Kob-Andersen mixture}

\begin{figure}
\includegraphics[width=8cm, height=8cm]{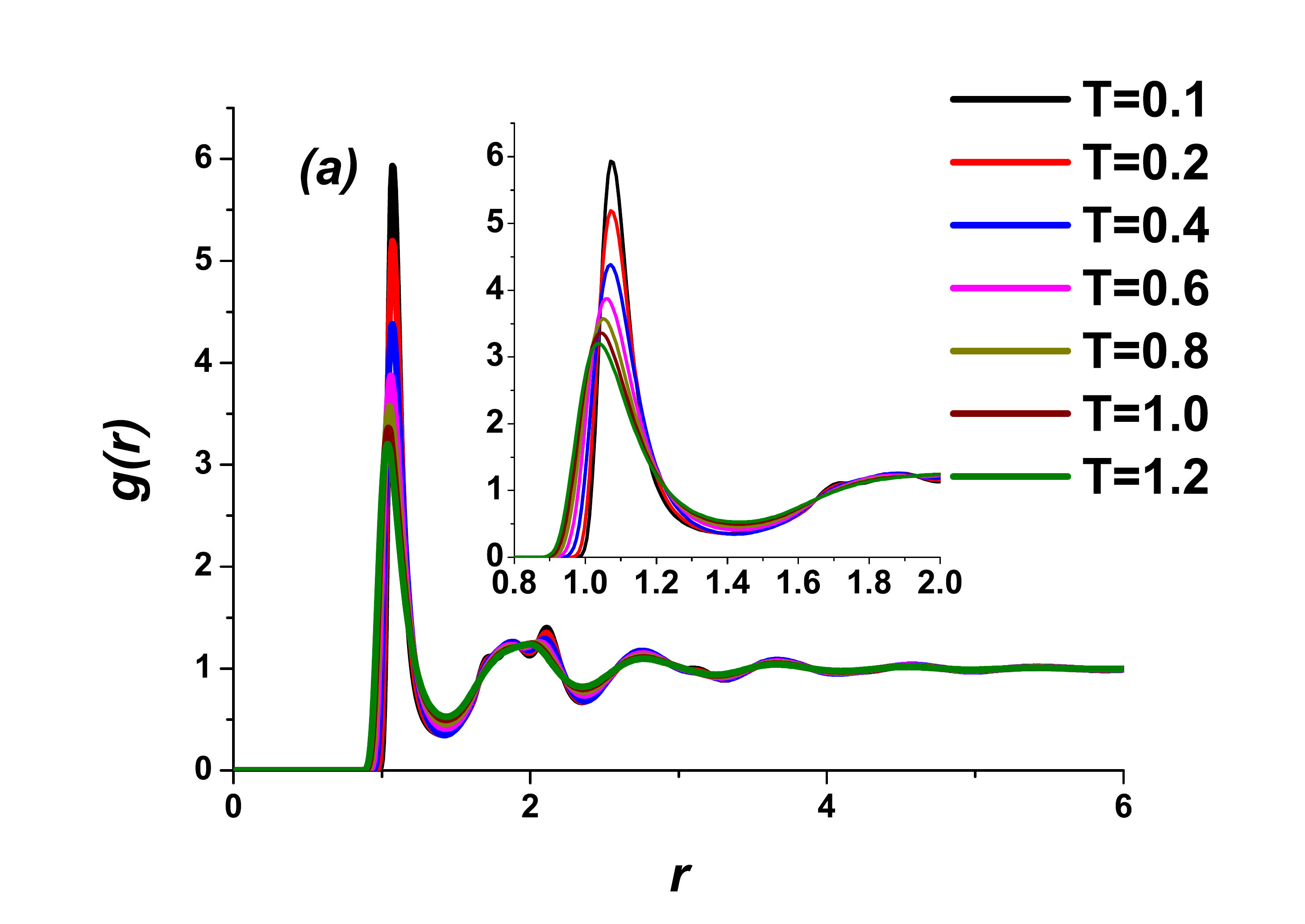}

\includegraphics[width=8cm, height=8cm]{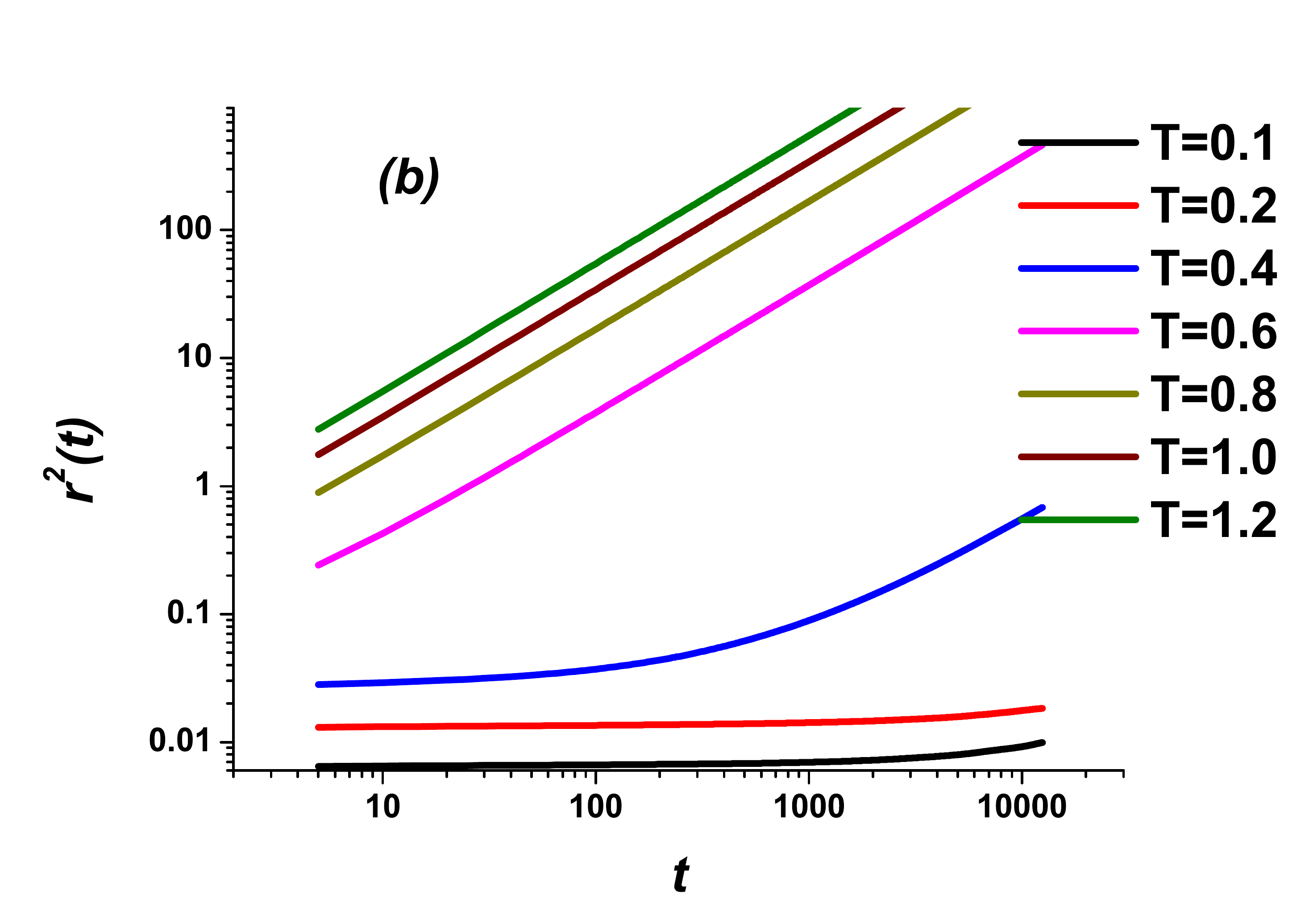}

\includegraphics[width=8cm, height=8cm]{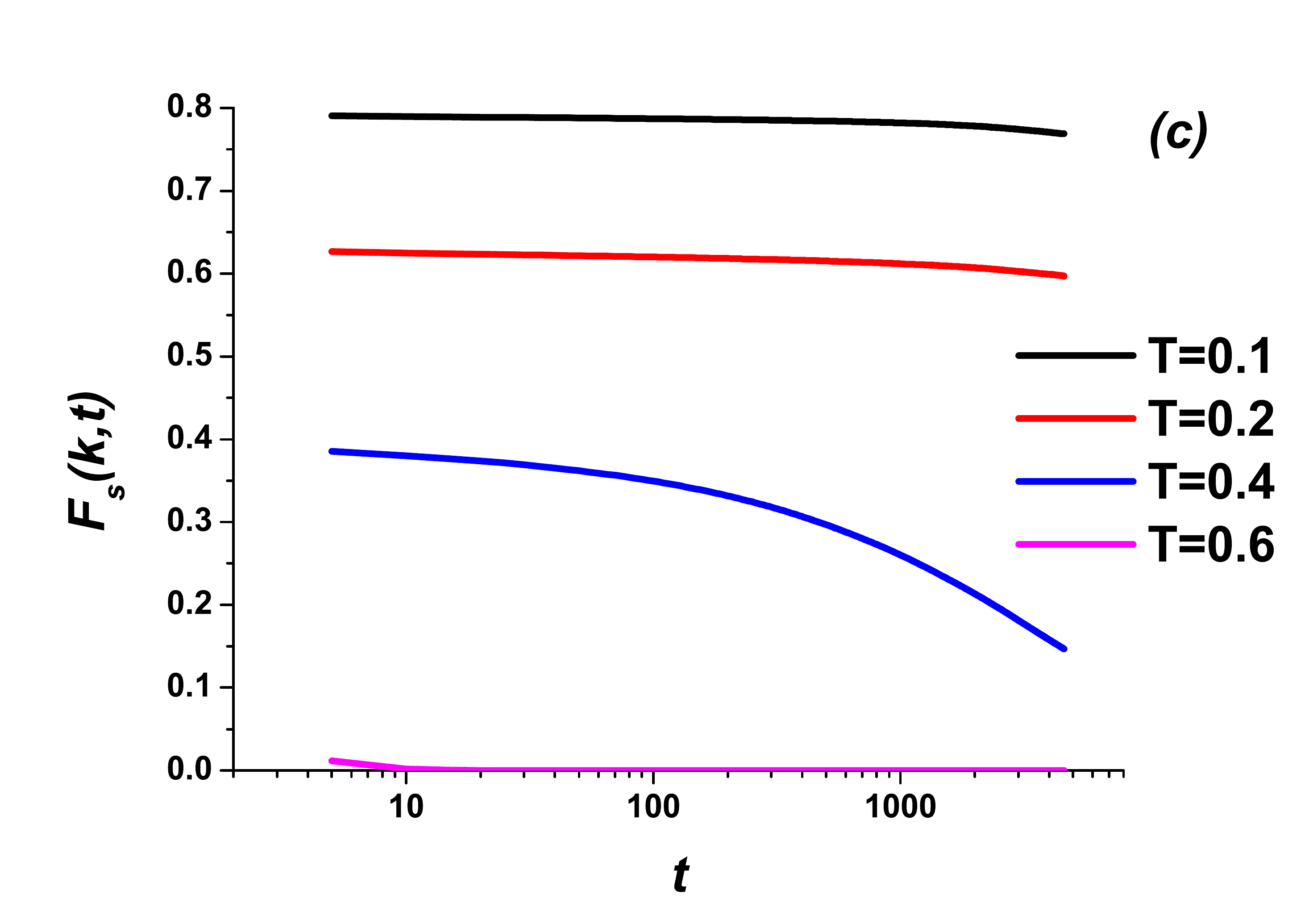}

\caption{\label{ka-app} (a) The radial distribution functions, (b)
mean square displacement and (c) intermediate scattering functions
of the A type of the KA mixture at $\rho=1.2$ and a set of
temperatures. The wave vector for the ISFs is $k=14.9$.}
\end{figure}

The radial distribution functions, mean square displacement and
intermediate scattering functions of the A type of the KA mixture
at $\rho=1.2$ and a set of temperatures are shown in Fig.
\ref{ka-app} (a)-(c).

\begin{thebibliography}{99}

\bibitem{gr1} C. A. Angell, Science {\bf 267}, 1924 (1995).

\bibitem{gr2} M. Ediger, C. Angell, and S. Nagel, J. Phys. Chem. {\bf 100}, 13200 (1996).

\bibitem{gr3} C. A. Angell, K. L. Ngai, G. B. McKenna, P. F. McMillan, and S. W.
Martin, J. Appl. Phys. {\bf 88}, 3113 (2000).

\bibitem{gr4} P. G. Debenedetti and F. H. Stillinger, Nature (London) {\bf 410}, 259 (2001).

\bibitem{gr5} V. Lubchenko and P. G.Wolynes, Annu. Rev. Phys. Chem. {\bf 58}, 235 (2007).

\bibitem{kobbinder} W. Kob and K. Binder,   {\it Glassy Materials and Disordered Solids: An Introduction to Their Statistical Mechanics}, World Scientific Publishing, 2005, Singapore.

\bibitem{gr6} L. Berthier, and G. Biroli, Rev. Mod. Phys. {\bf 83}, 587 (2011).

\bibitem{gr7} Jeppe C. Dyre, Rev Mod. Phys. {\bf 78}, 953 (2006).

\bibitem{gr8} G. Biroli1 and J.P. Garrahan, J. Chem. Phys. {\bf 138}, 12A301 (2013).

\bibitem{gr9} T.R. Kirkpatrick and D. Thirumalai, Rev. Mod. Phys. {\bf 87}, 183 (2015).

\bibitem{gr10} {\it Structural Glasses and Supercooled Liquids: Theory, Experiment, and Applications}, edited by P. G. Wolynes, and V. Lubchenko (Wiley, New York, 2012).

\bibitem{gr11} G. Parisi, P. Urbani, F. Zamponi, {\it THEORY OF SIMPLE GLASSES.
Exact Solutions in Infinite Dimensions}, (Cambridge University Press, 2020).

\bibitem{br} V.V. Brazhkin, JETP Letters  {\bf 112}, 745 (2020).

\bibitem{gg1} G. Biroli, J.-P. Bouchaud, F. Ladieu, J. Phys. Chem. B . {\bf 125} 7578-7586 (2021).

\bibitem{gg2} S. Albert et al Science {\bf 352}, 1308 (2016).

\bibitem{gg3} D. Chandler and J.P. Garrahan, Annu. Rev. Phys. Chem. {\bf 61}, 191 (2010).

\bibitem{gr12} M. D. Ediger, Annu. Rev. Phys. Chem. {\bf 51}, 99 (2000).

\bibitem{gr13} S. C. Glotzer, J. Non-Cryst. Solids {\bf 274}, 342 (2000).

\bibitem{gr14} H. C. Andersen, Proc. Natl. Acad. Sci. U.S.A. {\bf 102}, 6686 (2005).

\bibitem{gr15} S. Sastry, P. G. Debendetti, F. H. Stillinger, T. B. Schroder, J. C. Dyre, and S. C. Glotzer, Physica A {\bf 270}, 301 (1999).

\bibitem{gr16} T.R. Kirkpatrick, D. Thirumalai, Physical Review B {\bf 36}, 5388 (1987).

\bibitem{gr17} T.R. Kirkpatrick, D. Thirumalai, Physical Review A {\bf 37}, 4439 (1988).

\bibitem{gr18} T.R. Kirkpatrick, D. Thirumalai, P.G. Wolynes, Physical Review A {\bf 40}, 1045 (1989).

\bibitem{mct1} W. Gotze, Aspects of Structural Glass Transitions, in {\it Liquids, Freezing and the Glass Transition}, edited by J. P. Hansen, D. Levesque, and J. Zinn-Justin (Elsevier Science, Amsterdam, 1991), pp. 289-502.

\bibitem{mct2} W. Gotze and L. Sjogren, Relaxation Processes in Supercooled Liquids, Rep. Prog. Phys.
{\bf 55}, 241 (1992).

\bibitem{gr19} N. Petzold, B. Schmidtke, R. Kahlau, D. Bock, R. Meier, B. Micko, D. Kruk, and E. A.
Rossler,  J. Chem. Phys. {\bf 138}, 12A510 (2013).

\bibitem{gr20} D. Coslovich, M. Ozawa, and W. Kob, Eur. Phys. J. E {\bf 41}, 62 (2018).

\bibitem{glass88} H. Jonsson and H. C. Andersen, Phys. Rev. Lett. {\bf 60}, 2295 (1988).

\bibitem{dzugutov} M. Dzugutov, Phys. Rev. A {\bf 46}, R2984(R) (1992).

\bibitem{we1} Yu. D. Fomin, N. V. Gribova, V. N. Ryzhov, S. M. Stishov, and Daan Frenkel, J. Chem. Phys. {\bf 129}, 064512 (2008).

\bibitem{rysch} R. Ryltsev, N. Chtchelkatchev, and V. N. Ryzhov, Phys. Rev. Lett. {\bf 110}, 025701 (2013).

\bibitem{rysch1} R. Ryltsev, B. Klumov, and N. Chtchelkatchev, Soft Matter 11, 6991 (2015).

\bibitem{pd1} E. Zaccarelli, S. M. Liddle, and W. C. K. Poon, Soft Matter 11, 324-330 (2015).

\bibitem{pd2} S. Sarkar, R. Biswas, P. P. Ray, and B. Bagchi, J. Chem. Sci. Vol. 127, 1715 (2015)

\bibitem{pd3} Ya. Terada, Th. Keyes, J. Kim, and M. Tokuyama, AIP Conference Proceedings 1518, 776 (2013).

\bibitem{vink} R. L. C. Vink, G. T. Barkema, W. F. van der Weg, N. Mousseau, J. Non-Crysl. Solids 282, 248-255 (2001) .

\bibitem{ka1} W. Kob and H. C. Andersen, Phys. Rev. Lett. 73,  1376–1379 (1994).

\bibitem{ruocco1} G. Ruocco, M. Sampoli, and R. Vallauri, J. Chem. Phys. 96, 6167 (1992).

\bibitem{ruocco2} G. Ruocco, M. Sampoli, A. Torcini, and R. Vallauri, J. Chem. Phys. 99, 8095 (1993).

\bibitem{shih} J. P. Shih, S. Y. Sheu, and C. Y. Mou, J. Chem. Phys. 100, 2202 (1994).

\bibitem{abascal1} J. C. Gil Montoro, and J. L. F. Abascal, J. Phys. Chem. 97, 4211-4215 (1993)

\bibitem{abascal2} J. C. Gil Montoro, F. Bresme, and J. L. F. Abascal, J. Chem. Phys. 101, 10892 (1994).

\bibitem{jedl1} P. Jedlovszky, J. Chem. Phys. 111, 5975 (1999)

\bibitem{jedl2} P. Jedlovszky, J. Chem. Phys. 113, 9113 (2000).

\bibitem{lj-vor} T.-J. Hsu and Ch.-Yu. Mou, Mol. Phys. 75, 1329-1344 (1992).

\bibitem{voloshin} Vladimir P. Voloshin, Yuri I. Naberukhin, Nikolai N. Medvedev, and Mu Shik Jhon, J. Chem. Phys. 102, 4981 (1995).

\bibitem{si-vor} V. A. Luchnikov, N. N. Medvedev, A. Appelhagen, and A. Geiger, Mol. Phys. 88, 1337-1348 (1996)

\bibitem{ruff} A. Baranyai and I. Ruff, J. Chem. Phys. 85, 365 (1986).

\bibitem{qcgl} Yu. D. Fomin, Physics and Chemistry of Liquids, 58, 290-301 (2020).

\bibitem{lammps} S. Plimpton, J. Comp. Phys., 117, 1-19 (1995)

\bibitem{we1} Yu. D. Fomin, N. V. Gribova, V. N. Ryzhov, S. M. Stishov, and Daan Frenkel, J. Chem. Phys. 129, 064512 (2008).

\bibitem{rysch} R. Ryltsev, N. Chtchelkatchev, and V. N. Ryzhov, Phys. Rev. Lett. 110, 025701 (2013).

\bibitem{s1} N. V. Gribova, Yu. D. Fomin, D. Frenkel and V. N. Ryzhov, Phys. Rev. E 79, 051202 (2009).

\bibitem{s2} Yu. D. Fomin, V. N. Ryzhov, N. V. Gribova, Phys. Rev. E 81, 061201 (2010).

\bibitem{s3} Yu. D. Fomin, E. N. Tsiok, and V. N. Ryzhov, J. Chem. Phys. 135, 124512 (2011).

\bibitem{s4} Yu.D. Fomin, E.N. Tsiok, and V.N. Ryzhov, Eur. Phys. J. Special Topics 216, 165–173 (2013).

\bibitem{ka1} W. Kob, H. C. Andersen, Phys. Rev. Lett. {\bf 73}, 1376 (1994).

\bibitem{qcgl} Yu. D. Fomin, Physics and Chemistry of Liquids {\bf 58}, 290-301 (2020).

\bibitem{vink} R. L. C. Vink, G. T. Barkema, W. F. van der Weg, N. Mousseau, J. Non-Crysl. Solids 282, 248-255 (2001).

\end{thebibliography}
\end{document}